# Banning short-haul flights and investing in high-speed railways for a sustainable future?

Anne de Bortoli[1,2*], Adélaïde Feraille[3] :

**1** *LVMT, Ecole des Ponts ParisTech, University Gustave Eiffel*
**2** *CIRAIG, Chemical engineering department, Polytechnique Montreal*
**3** *Lab. Navier Laboratory, Ecole des Ponts ParisTech, University Gustave Eiffel*
\* Corresponding author; e-mail: anne.de-bortoli@enpc.fr

**Abstract**

Long-distance mobility sustainability, high-speed railways (HSR) decarbonization effect, and bans for short-haul flights are debated in Europe. Yet, holistic environmental assessments on these topics are scarce. A comparative life cycle assessment (LCA) was conducted on the Paris-Bordeaux transportation options in France: HSR, plane, coach, personal car, and carpooling. The overall ranking on four environmental indicators, from best to worst, is as follows: coach, HSR, carpooling, private car, and plane. Scenario analyses showed that increasing train occupancy decreases the environmental impact of the mode (-12%), while decreasing speed does not. Moreover, worldwide carbon footprints of electric HSR modes range 30-120 gCO2eq per passenger-kilometer traveled. Finally, a consequential LCA highlighted carbon paybacks of the HSR project. Under a business-as-usual trip substitution scenario, the HSR gets net-zero 60 years after construction. With a short-haul flight ban, it occurs after 10 years. This advocates for generalizing short-haul flight bans and investing in HSR infrastructure.

*Keywords:* Decarbonization investment; Carbon payback; High-speed rail; Short-haul flight ban; Long-distance mobility; Life Cycle Assessment;

# 1    Introduction

## 1.1    Context

Exploring the role of high-speed railway (HSR) development to make long-distance mobility more sustainable is not a new topic (Goodenough and Page, 1994; D'Alfonso et al., 2015), but has gained momentum recently. Indeed, HSR transportation can compete with air, conventional rail, and road modes, depending on socio-economic determinants (Bergantino and Madio, 2020). In terms of environmental sustainability, a new HSR generates environmental burdens for its construction, maintenance and end-of-life (EoL), but also impacts positively or negatively the environment due to its usage, depending on trip substitutions (Bergantino and Madio, 2020) and potential induced traffic (D'Alfonso et al., 2015). These multiple and variable effects tend to complexify the assessment of the role of HSR in a more sustainable future (D'Alfonso et al., 2015), despite most studies appraising environmental benefits from HSR developments (e.g., Sun et al., 2020; Yang et al., 2019; Chester and Horvath, 2012).

France carried out a major campaign to develop HSR (Zembri and Libourel, 2017), before the national Mobility 21 commission recommended to rather allocate public funds to existing railways and commuting (Commission Mobilité 21, 2013). The latest French HSR line started to be operated in 2017, but awareness around the environmental impacts of air transportation has revived the idea of expanding the HSR network in France (Girard, 2020). With the "LOM"—the latest Mobility Orientation Law in France (Legifrance, 2019), the French government aims to decrease air transportation, by banning (1) domestic flights for which an alternative train journey shorter than 2.5 hours exists as well as (2) airport extensions and construction. Banishing air transportation advertisement has also been suggested by the French Green deputies (Assemblée Nationale, 2019). These choices are motivated by fragmentary arguments: mainly the important contribution of air transportation to global greenhouse gas (GHG) emissions (estimated around 3.5% for direct emissions in 2011, based on updated radiative forcing factors (Lee et al., 2021)), and its high kilometer-based carbon footprint—145 gCO$_2$eq per passenger-kilometer traveled (pkt)—compared to bus and rail—respectively (resp.) 20 and 3.5 gCO$_2$eq/pkt according to the French amendment (Assemblée Nationale, 2019).

Nevertheless, while comparing the environmental impact of transportation modes is one of the first steps in guiding sustainable public policies, the previous figures are both biased and insufficient. Biased as they are based on the EN 16258 standards (AFNOR, 2013), which only takes into account the direct GHG emissions from transportation, *i.e.*, vehicle tailpipe emissions. However, depending on the mode, main contributions to the environmental performance of transportation modes can come from the rest of the vehicle life cycle (de Bortoli, 2021) or even the infrastructure (de Bortoli et al., 2017), especially for HSR modes (Chester and Horvath, 2012b). The insufficiency of the analysis also relates to the mono-criterium approach, considering climate change as the supreme environmental priority. Yet, to be reliable, transparent, and anticipate environmental burden-shifting, public policies must be based on a comprehensive environmental assessment of transportation modes. This means (1) including as many pertinent environmental dimensions as possible, and, (2) considering the vehicle and infrastructure impacts on their entire life cycle. These two requirements dismiss most of the classic environmental assessment methods used to support policy making, such as carbon footprinting, solely focused on climate change and sometimes just direct emissions (AFNOR, 2013), or the classical "environmental impact assessment" largely conducted by governments on transportation projects, mostly dedicated to assess local impacts due to the construction of new infrastructure, such as ecosystem fragmentation or habitat loss, without standardizing systematic assessment methods, and most of the time not considering the infrastructure impacts on the entire life cycle (Soria-Lara et al., 2020). Thus, the only suitable method to question the role of HSR, aviation, and other long-distance modes toward a sustainable carbon neutrality, because it addresses these two main criticisms around burden-shifting, is the "integrated modal life cycle assessment" (IMLCA), looking at various environmental impacts from both vehicles and infrastructure over their entire life cycle (de Bortoli and Christoforou, 2020).

Second, the environmental comparison of transportation modes even with an adequate method is not enough to understand the environmental role of HSR. Indeed, developing HSR requires the construction of new infrastructure, which generates environmental impacts, while other infrastructure already exists, may not have reached their capacity (roads, airports), and could be used instead. It means that, when comparing different modes using LCA, temporal aspects of the impacts are not considered, i.e., the fact that certain impacts have already occurred and cannot be avoided in the future—e.g., to produce the infrastructure or vehicles—while other impacts are to come. This leads to a bias in the analysis, as only future impacts matter to drive the environmental transition, *e.g.*, to reach net-zero. A consequential approach is therefore more appropriate in a nearer term to assess the sustainability impact of the development of HSRs, by comparing the environmental situation with and without the new infrastructure.

### 1.2 Research questions and objectives

This paper aims to answer the following questions: considering the urge to decarbonize our economy and reach a sustainable carbon neutrality by 2050 at the latest, is investing in HSR the wisest choices for the planet in terms of both public funds and environmental capital allocations? Should we ban short-haul flights? What are the best environmental alternatives for long-distance travel? First, section 2 reviews LCAs that address these questions. Thus, section 3 details our method - how attributional and consequential LCA models have been developed to answer these questions – on the Paris-Bordeaux link that falls under the short-haul flight ban announced by the French government. Then, section 4 presents and analyzes the results. Last, section 5 discusses (1) the seminal comparative US results in the light of our new models and results, (2) the adequate system boundaries for modal assessments on which to base unbiased environmental transportation recommendations, and (3) the options for sustainable long-distance travels toward net-zero.

## 2 Literature review

Many studies looked at the environmental impacts of HSR travel, compared HSR to other long-distance travel mode such as plane, car, coach, and sometimes considered the impacts of modal shifts toward HSR with a life cycle approach. Nevertheless, all of them present strong limitations that even question the validity of their conclusions.

## 2.1 Limitation 1 - Truncated LCA or heterogeneous system boundaries

The first strong limitation of the literature is an incomplete and/or heterogeneous life cycle considered in and within modes, biasing both modal comparisons and environmental impacts resulting from modal shifts.

Let's first look at attributional LCAs (ALCA), aiming at comparing the environmental performance of HSR, plane, and potentially other long-distance modes. Most studies considered the construction and operation stages, but excluded maintenance and disposal. Yue et al. excluded HSR infrastructure operation, maintenance, and disposal, as well as facilities used and material, energy and equipment transportation, due to a lack of data (Yue et al., 2015). Ha et al. only included operation stages, as well as infrastructure, rolling stocks, and aircraft construction, while excluding any EoL and maintenance of infrastructure and vehicles (Ha et al., 2011). Chang et al. assessed the environmental impacts of the Beijing-Shijiazhuang HSR line without any modal comparison with the same system boundaries (Chang et al., 2019).

More specifically, only a few studies considered the end-of-life (EoL) from the vehicles and infrastructure, and none of them consider recycling or reuse practices at the EoL for high-speed railway. This constitutes a major mistake, especially overestimating HSR travel impacts, as major environmental contributions of this mode come from the infrastructure, especially from the steel of the rail. Thus, appropriately considering the reuse and/or recycling of the rail – a most common practice worldwide – leads to a large decrease of the infrastructure impacts estimated (-21% in average for the Tours-Bordeaux HSR in France) (de Bortoli et al., 2020).

Secondly, all consequential LCAs (CLCA), used to calculate the environmental consequences of a new HSR under modal shifts, also presented heterogeneous system boundaries, besides systematically ignoring or truncating the EoL. The most incomplete type of system boundaries can be illustrated by Westin et al., who only considered the construction of the HSR line, then the negative emissions from modal shift related to the energy consumed by vehicles, i.e., the sole operation stage of competing modes (Westin and Kågeson, 2012). Chang and Kendall adopted similar system boundaries when estimating the carbon footprint from the San Francisco-Anaheim HSR project and its construction carbon payback (Chang and Kendall, 2011), just like the International Union of Railways (UIC, 2016). Quite a similar perimeter was considered by Wand and Sandberg, but adding the train construction impact, using environmentally-extended input output analysis (EEIO-A, sometimes called input-output (IO) LCA) (2012). Kortazar et al., and maybe Bueno et al. too (unclear calculations, also considered the sole direct negative emissions from modal shifts, but "only" excluded the EoL from the HSR mode assessment (Bueno et al., 2017; Kortazar et al., 2021). The system boundaries selected by Chen et al. when evaluating the Beijing-Shanghai HSR consequences are unclear, especially about maintenance stages, but infrastructure construction, as well as vehicle manufacturing and operation were accounted for (Chen et al., 2021). Akerman only excluded the manufacturing of aircraft when he calculated the climate change contribution due to modal shifts from air to the Europabanan Swedish HSR (Åkerman, 2011). The best consequential approaches considering system boundaries are those from Chester and Horvath, as well as Robertson, the first studies "only" excluding HSR EoL (Chester and Horvath, 2012a, 2010). Robertson declared considering complete life cycles, but used ecoinvent (EI) generic models that do not include EoL correctly as well, and only assessed the impact from the infrastructure life cycle due to material production, thus excluding earthwork, building machines and their consumption, in addition to roughly modeling the HSR maintenance stage as "40% of the construction stage" (Robertson, 2016).

## 2.2   Limitation 2 – data quality issues and potential inaccurate environmental assessments

The second strong limitation in the literature is the poor quality of the data used to assess the environmental impacts of the HSR and/or competing modes. The quality of input data - called life cycle inventories (LCI) in LCA - is crucial to reduce uncertainties and thus accurately assess environmental impacts of systems using LCA (Bamber et al., 2020). This data quality can be analyzed through the Pedigree matrix's 5 dimensions (Weidema et al., 2013; Weidema and Wesnæs, 1996): geographic, temporal, and other technological correlations, completeness, and reliability. Many studies do not address or realize the importance of data quality in correctly assessing the environmental impacts of HSR or its alternatives, simply gathering carbon footprints or emission factors of more or less similar systems from the literature (*e.g.,* Åkerman, 2011; Bueno et al., 2017; Chen et al., 2021; Kortazar et al., 2021). Most of the studies end up assessing their system with a critically low data quality. The consequence of this common practice is huge uncertainties on computed impacts, as the model does not represent the specific system studied, including specific materials, supply chains, construction techniques, and electricity mixes along the life cycle. For instance, Akerman did not use specific data for the Europabanan Swedish HSR construction assessment, but figures from the Bothnia line, non-transparently "adapted" (Åkerman, 2011). He also selected a theoretical energy requirement per seat for the HSR operation. Another example is Chen et al. (Chen et al., 2021) who used the $CO_2$ emissions factors of the 4 HSR evaluated by Baron et al. in a report from the French consulting company Systra (Baron et al., 2011): used in the context of Beijing-Shanghai, these carbon footprints originally calculated with French emission factors for two Asian and two French projects ends up with the worth score on geographical correlation (5/5, not representative), the score of 4/5 (second-worst score) in terms of temporal correlation (data more than 10 years old, less than 15 years old), and a probable worst score in terms of completeness (5: representativeness of data unknown). Chen et al. do not even provide the source of their data for airport construction. Kortazar et al. also evaluate the environmental impact of HSRs in Spain based on the carbon footprint from the HSR LCA by Tuchschmid et al. (2011) and generic emission factors for modal shift from EI v3.7 (Kortazar et al., 2021; Weidema et al., 2013). Finally, Bueno et al. also used highly derivative data. They recycled Baron et al.'s HSR emission factors, that obviously differ from the Basque Y line they assessed (worst score of 5 on the geographic correlation dimension). For instance, they roughly assumed bridge and tunnel lengths representing 70% of the line, while these lengths size the HSR construction impacts (Chang and Kendall, 2011). They also roughly estimated carbon savings from modal shifts through energy consumptions per pkt from the literature and conversion factors to obtain carbon emissions (Bueno et al., 2017). This study concludes on the detrimental impact of HSR in Spain in terms of carbon emissions, what could be explained by potential underestimated GHG savings from modal shifts, and overestimated HSR impacts due to a low 60-year lifespan, exclusion of the rail recycling benefits, and high rate of civil engineering structures on the line. This study is cited by several public policies, while the quality of the assessment may unfortunately raise doubts on the validity of its conclusions.

## 2.3 Limitation 3 – a restricted focus on carbon emissions

The third important limit of previously published environmental assessments of HSR development is the focus on $CO_2$ emissions (*e.g.*, Chen et al., 2021; Ha et al., 2011; Robertson, 2016) or GHG emissions ( *e.g.,* Åkerman, 2011; Chang and Kendall, 2011; Cornet et al., 2018), with sometimes potential confusions between the two (*e.g.*, Bueno et al., 2017). First, even if considering climate change is currently the most important environmental risk to mitigate, $CO_2$ only contributes to three thirds to GHG emissions according to Kyoto's protocol-compliant inventories (Ritchie and Roser, 2022). $CH_4$ and $N_2O$ contribute to resp. 17% and 6%m and must also be accounted for (Ritchie and Roser, 2022). Moreover, cutting-edge research on climate change assessment recommends considering all the gases with an impact on climate through the common Global Warming Potential 100 years (GWP100) indicator, as well as climate-carbon feedback for all climate forcers through the global temperature change potential 100 years (GTP 100) indicator (Jolliet et al., 2018). Second, the safe operating space for humanity is limited by ten identified planetary boundaries, including climate change, biodiversity loss, freshwater use, land-system change, ozone depletion, and ocean acidification (Steffen et al., 2015). Assessing the environmental pertinence of long-term development such as HSR deployment requires to include as many of these boundaries in their assessment. In addition to climate change, Chang et al. and Bueno et al. also considered an energy consumption indicator (Bueno et al., 2017; Chang et al., 2019), but it is also related to the climate change boundary or non-renewable resources. Chester and Horvath considered local pollutants (Chester and Horvath, 2012b, 2010), but did not link them to impacts, for instance on biodiversity, ozone depletion or other planetary boundaries. Yue et al. included more environmental dimensions, for instance non-renewable resource demand, acidification, and eutrophication, but did not assess the global damage to biodiversity, while LCA allows to do so (Yue et al., 2015). Cornet et al. proposed an interesting semi-quantitative comparison between GHG emissions and land transformation and occupation (LULUC) related biodiversity loss due to new HSR infrastructure (Cornet et al., 2018), but they forgot to put in perspective the effect of climate change on biodiversity loss, while the Millenium Ecosystem Assessment acknowledged climate change as one of the most important causes of the Sixth Extinction (Millennium Ecosystem Assessment, 2005). A more holistic assessment of the environmental impacts of HSR is thus required to provide a more complete insight of the environmental pertinence of HSR development.

## 2.4 Global limitations and good practices

To conclude, LCA – both attributional and consequential - has been used many times to try to compare HSR travel mode to alternatives or to assess the environmental consequences of a new HSR line, and especially due to modal shifts from flights. But most of the time, these studies used truncated or heterogeneous system boundaries, and generic impact factors or models from the literature, what are not sound practices to estimate accurately environmental impacts of specific projects. First, each stage of a transportation system's life cycle can be a major contributor to environmental impacts (see for instance Chester and Horvath, 2008; de Bortoli, 2021; de Bortoli et al., 2017). Moreover, the environmental impacts of similar objects are also naturally variable due to all the differences occurring along the life cycle, and these differences must be captured by tailored modeling of the most important contributors to environmental impacts. Additionally, climate change contributions or other impacts gathered in the literature are often calculated using different lists of climate forcers, and different characterization factors for these forcers (*e.g.,* the various GWP series from the Intergovernmental Panel on Climate Change (IPCC) work). Mixing quantification methods in an environmental assessment can generate non-transparency/difficult reproducibility, as well as increasing uncertainties, that must be resp. proscribed and minimized. Even studies using LCI background database and a proper LCA software to compute impacts consistently often mix the background databases, such as Chester and Horvath, mixing EI v1.3 and EEIO inventories (Chester and Horvath, 2012b, 2010, 2008), Yue et al. using the Chinese Core Life Cycle Database (CLCD) complemented by EI 3.0 (Yue et al., 2015), or Chang et al., mixing the CLCD and Chinese EEIO inventories (Chang et al., 2019). In particular, EI has a wide climate forcer coverage, while EEIO databases are built on national inventories that only monitor a few GHG, ending in distortion in the results of climate change contributions. Finally, no LCA addressed the environmental impacts of HSR under short-haul flights bans.

# 3  Method

We aim at using both A- and C-LCA to answer our research questions with a high-quality environmental assessment method, including consistent system boundaries, background database, and life cycle impact assessment method, as well as high-quality primary data.

## 3.1  Case study

Train is commonly considered as an environmentally friendly transportation mode and is often subsidized by governments in Europe (Autorité de régulation des transports, 2020). Europe has an important railway network, France having the second-longest network with a total of 28100 km (Autorité de régulation des transports, 2020). From the beginning of rail mobility in France in 1841, the passenger rail traffic intensity has regularly increased—and doubled these last 40 years—to reach 1238 million travelers and 91.5 billion pkt annually in 2018 (SNCF, 2019). The massive HSR deployment plan from the last decades must have contributed to this surge. In 2019, 2600 km of HSR were operated in France (Autorité de régulation des transports, 2020), and the newest 302 km-long HSR route runs between the cities of Tours and Bordeaux since July 2017 (de Bortoli et al., 2020), reducing passenger time between Paris and Bordeaux by 30%. This new service is now in competition with other modes on the Paris-Bordeaux corridor, including short-haul flights and road modes—coach, personal car, and carpooling. The HSR line should also be affected by the short-haul flight ban announced by the French government, as it offers a 2-hour-long alternative to the air link. The objective of the case study is to compare the environmental impacts of the different modes on the Paris-Bordeaux corridor, to explore the sensitivity of these impacts to different operational parameters, and to assess the environmental impact of the HSR project from a consequential point of view, including carbon payback times, without and with the short-haul flight ban. This line is also chosen because of our access to primary data, *i.e.*, specific data on the life cycle of the infrastructures and vehicles, for each transportation alternative assessed, for example ex-post HSR construction spreadsheets and maintenance scheme, or traffic data. Primary data are necessary to accurately assess the transportation options.

## 3.2 Calculation overview

To assess the environmental relevance of the Paris-Bordeaux HSR project—and of HSR in general, we will conduct a multicriteria environmental assessment of the competing modes on the Paris-Bordeaux corridor using the LCA methodology to compare their impacts. LCA is a standardized method to assess the environmental impact of a system—a service, a product, an organization—on various quantitative dimensions and potentially over the entire life cycle of the system (International Organization for Standardization, 2006a, 2006b). To compare the competing modes, only mono-modal trips will be considered. For instance, the aerial alternative will be considered from airport to airport, neglecting the trips to access and leave the hub. These additional trips can be done using public transportation: by RER train, the light urban train in Paris that has a very low environmental impact in Paris (de Bortoli and Christoforou, 2020), and by bus in Bordeaux. Thus, we assume this would not substantially change the comparison. A contribution analysis will be carried out to understand the main drivers of the environmental performance of each mode depending on the indicator considered.

Then, scenario analyses (SAs) will be conducted to highlight key parameters of the four environmental impacts of the transportation modes assessed, and to understand their influence. Technological progress is often put forward as the preeminent means of solving the environmental problems of our time. To test this hypothesis in the case of long-distance journeys, we will calculate the environmental performance of trains, planes, and cars with conventional engines and more advanced engines. Then, we will perform sensitivity analyses focused on the HSR, by testing the impact of the electricity mix (selection of a set of geographical areas with electricity mixes covering in the worldwide carbon footprint range), the train capacity, and the commercial speed. This sensitivity analysis also aims at showing the natural variability of environmental impacts under different geographic or technological circumstances, to make the community aware of the inaccuracies related to selecting erratically emission factors from the literature, and the importance of performing a consistent tailored LCA when assessing specific transportation systems.

Finally, consequential LCAs will also be carried out to quantify the net carbon footprint of the HSR project, considering the infrastructure life cycle and the impact of trip substitutions from air to rail under two scenarios: a business-as-usual (BAU) scenario and a scenario where short-haul flights are banned. The new HSR was opened mid-2017. Based on traffic data from the French aviation directorate (DGAC, 2020, 2019, 2018), we calculated that the annual air traffic of the link decreased by 7.1% in 2017, 17.3% in 2018, and 3% in 2019, showing a clear reduction trend in the number of passengers (see supplementary material). Nevertheless, it is still difficult to draw a precise trajectory on the long-term annual modal shifts due to the new HSR service, especially because of exogenous factors (e.g., national strike in 2019–2020). Moreover, a decrease in the number of passengers does not mean a decrease in the number of flights—and thus a decrease in environmental impacts—as it depends on the airline company strategy. Under the BAU scenario, we assumed a steady annual modal shift from air to rail at around 250 000 passengers per year (average between the rail operator and the plane operator figures), i.e., a drop of 4 return flights daily over the 16 operated before the HSR opening (Provenzano, 2019). Considering the respective distances between Paris and Bordeaux by plane (507 km) and by rail (497 km), and the carbon footprint per pkt with or without the infrastructure respectively by air and rail, we will estimate the carbon payback period of the HSR project (see calculation details in the supplementary material). The second scenario—referred to as the "ban scenario"—accounts for 12000 flights avoided per year, i.e., 16 return flights a day based on the daily average operation of the line in 2021, equal to 1.6 million of passenger-trip traveled (Mptt) avoided annually.

### 3.3 Scope and functional units

The first part of the study relies on a process-based attributional LCA of transportation modes, considering both the infrastructure and the vehicles, as well as the entire life cycle—from the production to the End-of-Life (EoL). This approach is defined as an "integrated modal LCA" (de Bortoli, 2021). The system boundaries of such a method are presented in Figure 1. Two functional units (FU) will be considered for the comparative ALCAs: "moving one person from Paris to Bordeaux", i.e., the ptt, and "moving one person over one kilometer", i.e., the pkt.

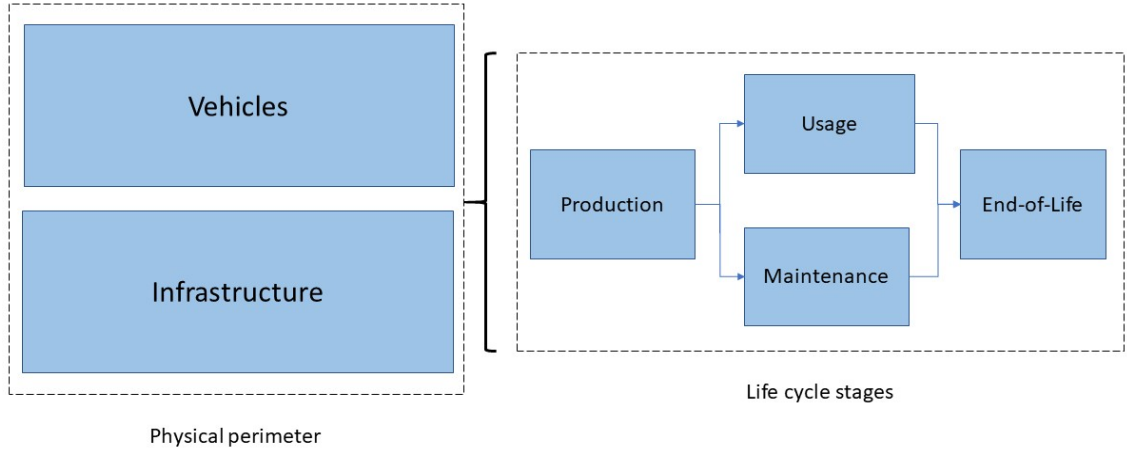

**Figure 1 System boundaries of the integrated modal LCA method**

The consequential approach will be based on a consequential LCA (see for instance Weidema et al., 2018), and will quantify the net GHG emissions of the HSR project, considering the impact of the traffic on the HSR and the avoided emissions due to traffic displacement from aerial transportation, again based on integrated LCA.

### 3.4 Integrated modal LCA equation

The LCA software used is OpenLCA 1.9. The formula to calculate the environmental impact of a mode $i$ per pkt using integrated modal LCA is presented in Equation (1) (de Bortoli and Christoforou, 2020), with $EF_{veh,i}$ the environmental factor from the vehicle component of the mode $i$ per pkt, $EF_{infra,i}$ the environmental factor from the infrastructure component of the mode $i$ per pkt, $EF_{1veh,i}$ the environmental factor of one average vehicle of the mode $i$ over its life cycle, $PKT_{1veh,i}$ the lifetime mileage of one vehicle, $PKT_{ij}$ the number of pkt on the infrastructure $j$ for the mode $i$, $a_{i,j}$ the "specific infrastructural demand" (Spielmann et al., 2007) of type $j$ by the mode $i$ (see further explanations below), $q_j$ the number of units of the infrastructure $j$ used by the mode $i$, $EF_{1u,j}$ the environmental factor of one unit (surface or length) of the infrastructure $j$, $b_{ij}$ the infrastructural allocation factor of the infrastructure $j$ for the mode $i$, and $VKT_{ij}$ the number of vehicle-kilometers traveled by mode $i$ on the infrastructure $j$.

$$EF_i = EF_{veh,i} + EF_{infra,i} = \frac{EF_{1veh,i}}{PKT_{1veh,i}} + \sum_j \frac{1}{PKT_{i,j}} \cdot a_{ij} \cdot q_j \cdot EF_{1u,j} = \frac{EF_{1veh,i}}{PKT_{1veh,i}} +$$

$$\sum_j \frac{1}{PKT_{i,j}} \cdot \frac{b_{ij}.VKT_{ij}}{\sum_i b_{ij}.VKT_{ij}} \cdot q_j . EF_{1u,j} \quad (1)$$

We multiply $EF_i$ by the number of kilometers of the trip to obtain the environmental impact per ptt.

### 3.5 Characterization methods

Our set of indicators is based on the v1.47 IMPACTWorld+ (IW+) characterization methods (Bulle et al., 2019), the Cumulative Energy Demand (CED), as well as the IPCC 2013 GWP. A set of four indicators, encompassing primary energy consumption (addition of the sub-indicators of the CED method), damages to ecosystems and human health (IW+ method), as well as climate change contribution, are selected. We choose the endpoint level of IW+ indicators as the objective of this LCA is to provide a comprehensive snapshot of the environmental sustainability of long-distance travel modes for decision makers, considering as exhaustively as possible the complexity of environmental issues. Endpoint indicators, sometimes considered as less reliable than midpoint indicators as they are based on more complex causality chains calculations, have the advantage of widely considering the different quantifiable environmental dimensions in a restricted number of indicators, without using subjective weighting methods. For instance, the indicator of damage to ecosystems aggregates the various impacts of the transportation options on biodiversity due to short- and long-term climate change effects, ocean and other acidifications, eutrophication, ecotoxic releases, water thermal pollution and scarcity, as well as LULUC, all along the life cycle of the transportation systems. The two midpoint indicators—primary energy consumption and contribution to climate change—are calculated for comparison purposes as they are the most popular in public policies and scientific publications.

### 3.6 Modeling key parameters

The main characteristics of the parameters used in our transportation LCA models are presented in Table 1, as these "key parameters" (Jolliet et al., 2005) size the environmental impacts of the modes and must thus be presented transparently. They will be detailed further in the article and its supplementary material (text document and Excel spreadsheets). The capacity of the vehicles – *i.e.*, maximum number of people in the vehicles - is expressed in numbers of passengers (pax).

**Table 1 Synthesis of the modeling main parameters**

| Mode (+ distance) | Infrastructure lifespan (years) | Vehicle type | Weight (t) | Vehicle lifetime mileage (km) | Capacity (pax) | Occupancy rate (%) |
|---|---|---|---|---|---|---|
| Train (497 km) | Rail: 120 | Atlantique | 445 | 12 000 000 | 454 | 68% |
| | Stations: 100 | Avelia Euroduplex Oceane | 401 | 12 000 000 | 556 | 68% |
| | | Ouigo | 401 | 12 000 000 | 634 | 76% |
| Plane (507 km) | Airport: 100 | A320-ceo | 43 | 28 871 000 | 155 | 86% |
| | | A320-neo | 44 | 28 871 000 | 165 | 81% |
| Road (589 km) | Earthworks: 100 | Coach | 13 | 1 000 000 | 51 | 69% |
| | (Sub)base course: 30 | Diesel car | 1.3 | 200 000 | 5 | 44% |
| | Wearing course: 13 | Petrol car | 1.3 | 200 000 | 5 | 44% |
| | | Diesel carpooling | 1.3 | 200 000 | 5 | 68% |
| | | Petrol carpooling | 1.3 | 200 000 | 5 | 68% |

### 3.7 Life cycle inventories

Ecoinvent v3.2, with the cut-off allocation system, is used as a background database. EI is the most comprehensive Life Cycle Inventories (LCIs) database, containing international industrial data enabling to assess the environmental impact of energies, resource extraction, chemical products, metals, agriculture, waste management services and transportation services (Wernet et al., 2016). The foreground LCIs have been collected from transportation operators, infrastructure construction companies and concessionaires, and completed by data from the literature when needed. They are summarized below by mode, and exhaustively explained in the supplementary material. EoL allocation is a 100:100 approach, considering a credit for avoided virgin production at a rate of 100% (Allacker et al., 2017).

### 3.7.1 High-speed rail modes

#### 3.7.1.1 Railways and stations

LCIs of the railway are adapted from the study of de Bortoli et al. (2020) (see supplementary material). The HSR section only goes from Poitiers to Bordeaux, over 302 km, and the beginning of the line is a standard railway. A theoretical full HSR line has been modeled by scaling up the real HSR section LCI. Because no other data were available to conduct the LCA of the HSR stations, we estimated the type of building and the areas of the four other HSR stations using Google Maps and its different tools (measure a distance, Satellite and Street View). The characteristics of the stations are detailed in the supplementary material.

#### 3.7.1.2 Trains

Two kinds of carriages are considered on the HSR corridor: the Atlantique train that is a single-stage train and the Avelia Euroduplex Oceane trains that is a double-stage train (called "Duplex train" in the rest of the article). New LCIs are developed, as the only model in EI is a German internal combustion engine (ICE) single-stage train. Details on the LCIs can be found in the supplementary material. For the production stage, material quantities were provided by Alstom's experts, while the rest of the inventories were compiled from the existing EI LCI on the production train process, which also includes the EoL (Spielmann et al., 2007). Transportation activities between the diverse manufacturing sites to the final assembling plant in Belfort, France, are detailed in the supplementary material. The locations of the different sites come from declarations from the French Ministry of Transportation (Ville, Rail et Transports, 2019). The average electricity consumption on the HSR section has been measured equal to 22.9 kWh/km for Euroduplex Oceane trains and 27.1 kWh/km for Atlantique trains by the concessionaire of the line, LISEA, for maximum speeds ranging between 300 km/h and 320 km/h. The maintenance has been modeled using the EI's process "*maintenance, train, passenger, high-speed, DE*", rescaled based on the train's weights (ratio 401 tons for the Euroduplex Oceane train / 640 tons for the German train). Finally, the EoL scenario is based on a report from the French Professional syndicate for recycling companies and includes recycling, landfill, and other disposals (FEDEREC and ADEME, 2017). A distance of 400 km has been considered from the dismantling point to the recycling platform (FEDEREC and ADEME, 2017).

### 3.7.2 Plane mode

#### 3.7.2.1 Airports

Three airports are used to travel by plane between Paris and Bordeaux: two in the French capital—Paris-Charles de Gaulle (CDG) airport which represents 40% of the Paris-Bordeaux flight departures based on our calculations, and Orly airport, which represents the remaining 60% (DGAC, 2014)—and the one in Bordeaux-Mérignac. Because airports did not share any data to conduct an LCA, the process "Airport Construction—RER" from the EI database was reused. As the data change significantly from one airport to another depending on the size, the construction techniques, and the electricity mix, the EI process was adjusted for each of the three French airports considered using a method detailed in the supplementary material.

#### 3.7.2.2 Aircraft

The air link between Paris and Bordeaux is operated by Air France airline. The fleet comprises planes from the Airbus A320 family. Two A320 models are used: the classical A320ceo ("ceo" standing for Current Engine Option) and the new model A320neo ("neo" standing for New Engine Option). The construction and EoL LCIs are built on the results of the PAMELA project of Airbus, an LCA conducted on an A330-200 aircraft (de Oliveira Fernandes Lopes, 2010) and adapted to the specific aircraft used on the link Paris-Bordeaux. Three types of components are modeled: the aircraft frame, the seats, and the engines. Details can be found in the supplementary material. The aircraft maintenance is out of the scope, due to a lack of data. The use stage has been modeled particularly carefully as it may carry most of the impact (Chester, 2008). Two use stage cycles were considered: the Landing and Take-Off cycle (LTO) and the climb/cruise/descent (CCD) cycle. The duration of each period of each cycle is set based on Chester and Horvath data (Chester and Horvath, 2008). Kerosene consumption and emissions for the CDD cycle were calculated using the "Master emissions calculator 2016" spreadsheet (EMEP/EEA, 2017a) accompanying the EMEP/EEA air pollutant emission inventory guidebook 2016 (European Environment Agency, 2016). For the LTO cycle, the dedicated LTO emissions calculator was used, specifying the type of aircraft, the type of engine, as well as the specific airport and year considered (EMEP/EEA, 2017b). The EMEP spreadsheet does not allow to specify the type of engine: a 12% improvement in consumption and emissions has been considered, accounting for the LEAP-1A26 engine performance and the extra-weight of the A320neo (Hensey and Magdalina, 2018).

### 3.7.3 Road modes

Three road transportation modes are studied: average passenger car transportation, using gasoline or diesel, carpooling, and coach. These vehicles travel on the highway A10, linking Paris and Bordeaux.

#### 3.7.3.1 Cars and coaches

Standard vehicle categories are created to build the related LCIs. The average weight of a passenger car in France is 1278 kg (Compte, 2018). The representative coach is set as a Volvo B10M / Berkhof Axial 50, a two-axle bus weighing 13 300 kg with a capacity of 51 passengers (UK government, 2016). The material compositions of the average vehicles come from EI. The EoL treatment is included in the EI production process for vehicles. The maintenance of both passenger cars and coaches are modeled directly using the existing EI processes, namely "*maintenance, passenger car—RER*" and "*market for maintenance, bus—GLO*". For the use stage, the emission model selected is the European EMEP/EEA guidebook. The "EMEP guidebook" gives emission factors of different types of vehicles for different speeds depending on a set of parameters like the gradient of the road and the occupancy rate of the coach to consider the impact of load on consumption and emissions. Between Paris and Bordeaux, speeds on the highway A10 are limited to 130 km/h for light vehicles and 90 km/h for heavy vehicles, while average speeds for these vehicles are respectively 118 km/h and 88 km/h (73-55 mph) (de Bortoli et al., 2022). The coach consumes diesel, while the passenger car can consume either diesel or petrol (=gasoline).

*3.7.3.2 Roadways*

With 62.9 million trucks traveling over the highway A10 each year, it is a "class TC7" road according to the French catalog of road structures (Corte et al., 1998), with a PF3-class platform (George et al., 2001). We modeled the cross-section and construction materials requirement for construction using the French road catalog (Corte et al., 1998), following the French specifications for highway subbase and base courses (LCPC-Sétra, 1998). Details can be found in the supplementary material. The maintenance consists of milling and rebuilding a very thin asphalt concrete layer over 2.5 cm after thirteen years, the base after thirty years, and the earthworks after 100 years, based on a French survey on pavement layer lifespans (de Bortoli, 2018).

# 4 Results and interpretation

## 4.1 ALCA: modal impacts and comparisons

In this section, we present the environmental impacts and rankings of the different means of transportation on the four selected indicators, depending on the two functional units previously proposed (pkt and ptt).

### 4.1.1 Absolute results

Table 2 gives the environmental impacts of each means of transportation by ptt and pkt. These results will be compared first per ptt in section 3.1.2, and second per pkt in section 3.1.3, to test the importance of the functional unit on the ranking of long-distance transportation modes. When considering the vehicle and infrastructure life cycles, traveling by rail onboard of a Duplex or an Atlantique train resp. emits 37 or 38 gCO$_2$eq/pkt, while it emits only 21 g by coach, 69 or 80 g when carpooling (resp. in diesel and gasoline cars), 106 or 123 gCO$_2$e when using private car (resp. diesel and gasoline), and between 213 and 242 g to fly (resp. with CEO and NEO aircraft).

**Table 2 Environmental impacts of each means of transportation per functional unit (pkt and ptt)**

| Mode | FU | Climate change (kgCO$_2$eq) | Ecosystem quality (PDF.m$^2$) | Human health (DALY) | Energy (MJ) |
|---|---|---|---|---|---|
| **Railway, Duplex** | pkt | 3.65 10$^{-2}$ | 4.24 10$^{-2}$ | 1.09 10$^{-6}$ | 5.84 10$^{-1}$ |
|  | ptt | 1.81 10$^{1}$ | 2.11 10$^{1}$ | 5.41 10$^{-4}$ | 2.90 10$^{2}$ |
| **Railway, Atlantique** | pkt | 3.80 10$^{-2}$ | 4.40 10$^{-2}$ | 1.12 10$^{-6}$ | 7.47 10$^{-1}$ |
|  | ptt | 1.89 10$^{1}$ | 2.19 10$^{1}$ | 5.59 10$^{-4}$ | 3.71 10$^{2}$ |
| **Air, ceo** | pkt | 2.42 10$^{-1}$ | 2.58 10$^{-1}$ | 4.19 10$^{-6}$ | 3.91 10$^{0}$ |
|  | ptt | 1.23 10$^{2}$ | 1.30 10$^{2}$ | 2.12 10$^{-3}$ | 1.98 10$^{3}$ |
| **Air, neo** | pkt | 2.13 10$^{-1}$ | 1.96 10$^{-1}$ | 3.99 10$^{-6}$ | 3.43 10$^{0}$ |
|  | ptt | 1.08 10$^{2}$ | 9.93 10$^{1}$ | 2.02 10$^{-3}$ | 1.74 10$^{3}$ |
| **Car, diesel** | pkt | 1.06 10$^{-1}$ | 1.09 10$^{-1}$ | 2.20 10$^{-6}$ | 1.62 10$^{0}$ |
|  | ptt | 7.27 10$^{1}$ | 7.46 10$^{1}$ | 1.40 10$^{-3}$ | 1.10 10$^{3}$ |
| **Car, gasoline** | pkt | 1.23 10$^{-1}$ | 1.27 10$^{-1}$ | 2.37 10$^{-6}$ | 1.87 10$^{0}$ |
|  | ptt | 6.26 10$^{1}$ | 6.42 10$^{1}$ | 1.30 10$^{-3}$ | 9.56 10$^{2}$ |
| **Carpooling, gasoline** | pkt | 7.99 10$^{-2}$ | 8.20 10$^{-2}$ | 1.54 10$^{-6}$ | 1.21 10$^{0}$ |
|  | ptt | 4.71 10$^{1}$ | 4.83 10$^{1}$ | 9.05 10$^{-4}$ | 7.14 10$^{2}$ |
| **Carpooling, diesel** | pkt | 6.88 10$^{-2}$ | 7.06 10$^{-2}$ | 1.42 10$^{-6}$ | 1.05 10$^{0}$ |
|  | ptt | 4.05 10$^{1}$ | 4.16 10$^{1}$ | 8.38 10$^{-4}$ | 6.18 10$^{2}$ |
| **Coach** | pkt | 2.07 10$^{-2}$ | 2.15 10$^{-2}$ | 5.43 10$^{-7}$ | 3.13 10$^{-1}$ |
|  | ptt | 1.22 10$^{1}$ | 1.27 10$^{1}$ | 3.20 10$^{-4}$ | 1.85 10$^{2}$ |

4.1.2 Comparison per passenger-trip traveled

As shown in Table 2 and Figure 2, when considering the ptt as a functional unit, the ranking globally indicates that traveling by plane is the most impacting mode on all indicators, followed by traveling by private car, carpooling, high-speed train, and finally coach. The results also show the importance of the multi-criteria approach as the ranking on different metrics can vary: here, the two technologies of planes give very similar impacts in terms of human health damage, while the most recent technology reduces by 10% to more than 20% the impacts on the three other metrics.

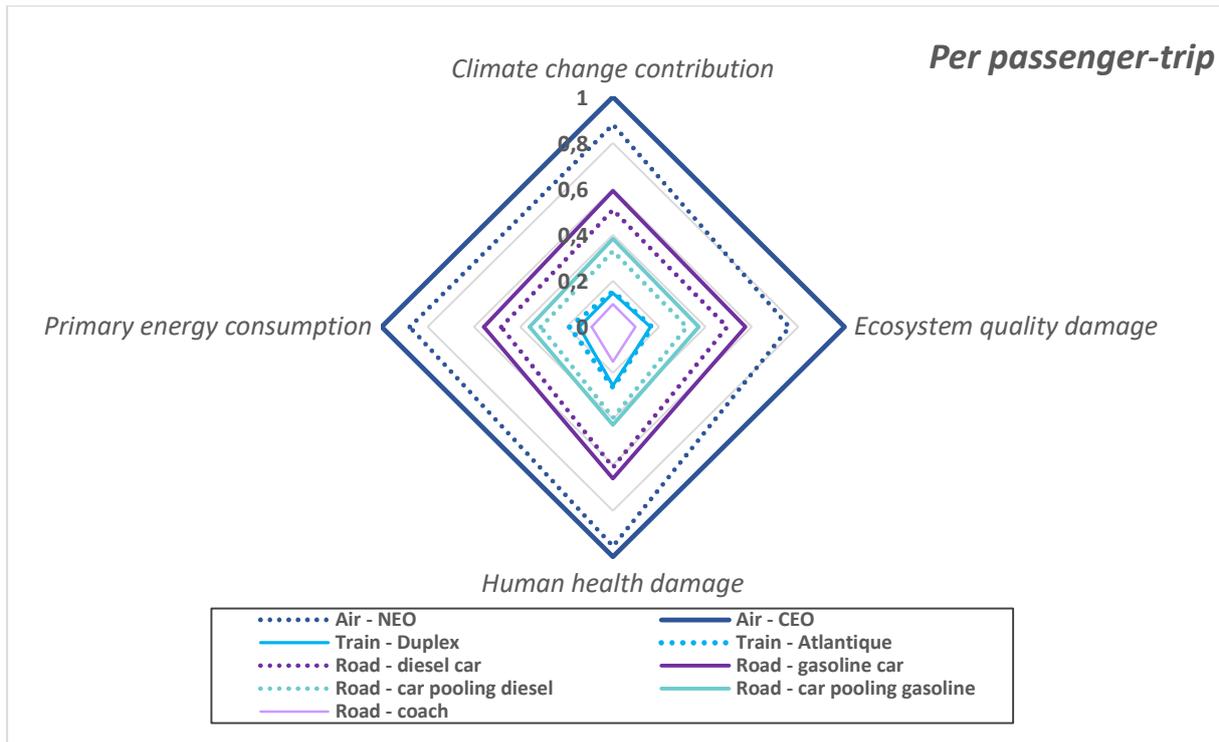

**Figure 2 Normalized impacts of the different means of transportation per passenger-trip traveled**

4.1.3   Per passenger-kilometer traveled

The standard functional unit to compare the environmental impact of transportation modes is the pkt. However, a trip between two cities presents different distances depending on the mean of transportation used. We want to check if the distance difference changes the environmental ranking of the modes on the Paris-Bordeaux case study. While Figure 2 presented the ranking based on the ptt, Figure 3 provides the ranking of the transportation modes studied per pkt. Although the ranking remains globally stable, the environmental performance of road modes gets better per pkt than per ptt respectively to other modes, and the train performance deteriorates slightly. This is explained by the distances 16% longer by road than by plane, and 19% longer by road than by train. Indeed, Table 1 shows that traveling from Paris to Bordeaux is shorter by train (497 km), followed by plane (507 km) then by coach and car (589 km). Thus, a trip using the roads "consumes" more kilometers than using the railway, explaining the distortion in the two figures. With this FU, coach appears better for the environment than HSR (-43 to 50% depending on the metric), compared to the difference with the ptt as the FU (-33 to -41%).

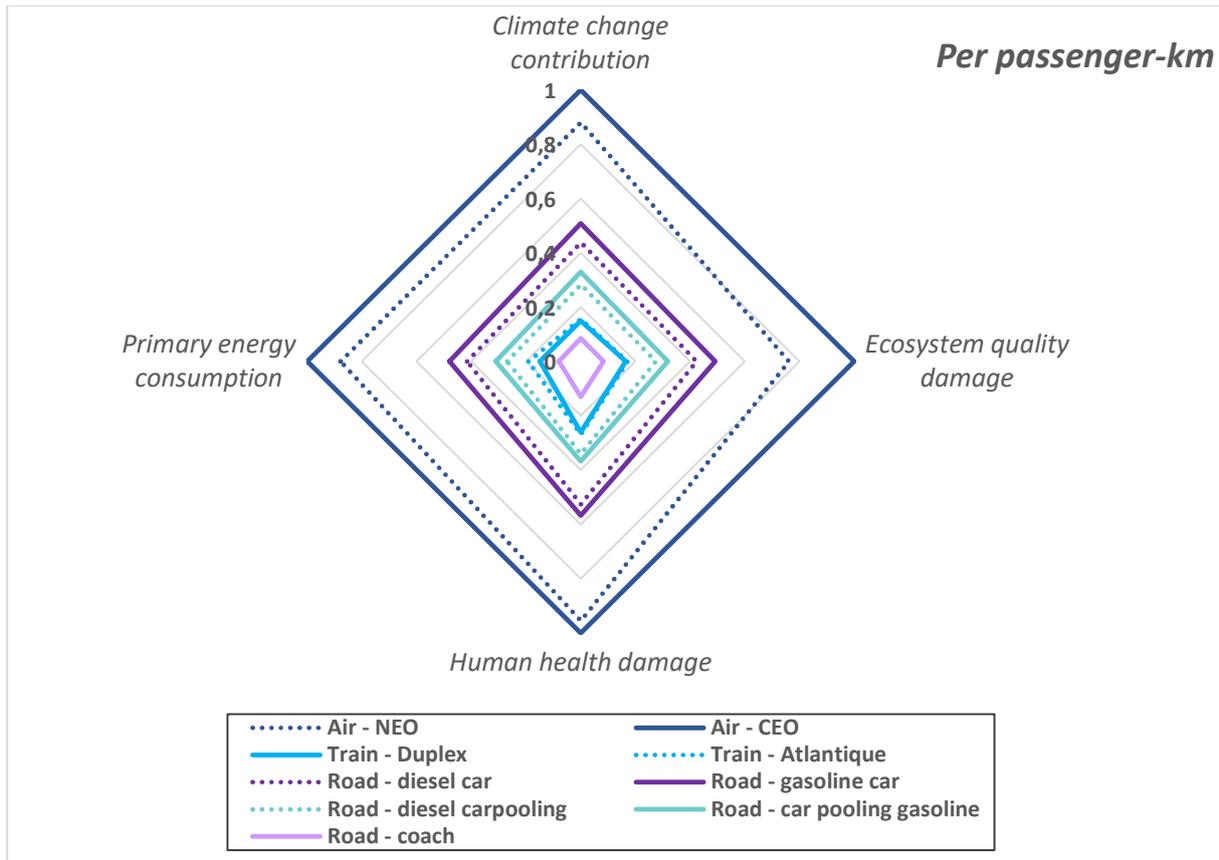

**Figure 3 Normalized impacts of the different means of transportation per passenger-kilometer traveled**

## 4.2 ALCA: contribution analyses

Figure 4 to Figure 7 show the distribution of the environment contributions from the different components of the transportation modes. The different contributions presented are the vehicle use stage (tagged as "use", in blue), the rest of the vehicle life cycle (manufacturing amortization, vehicle maintenance, and EoL, in magenta), and the infrastructure life cycle (raw material extraction, construction, maintenance, EoL, in gray). For plane and road transportation, the use stage globally brings most of the impacts (from 30 to 80% of the total impact for air; from 25 to 80% for car), while for train transportation, the most important contribution is mostly due to the infrastructure (between 25 and 90% from the railway). The contribution of the different components of each transportation mode will be described further below.

4.2.1 Train travel

In Figure 4, we can see that the railway represents the main contributor to climate change, ecosystem quality damage, and human health damage for the train mode, while the second contributor is the use stage. For primary energy consumption, the trend is different: the use stage is the main contributor, and the railway is the second most important contributor. This is due to the major contribution of imported steel to build the rail to climate change and damage to ecosystem and human health on one side, and to the major contribution of the French electricity consumed in the use stage on the other side, that represents a high primary energy consumption, while it has a low carbon intensity as it mostly relies on nuclear energy and hydrogeneration. The vehicle contribution is negligible—between 1 and 3% of the first two impacts but represents resp. 9 and 5% of damage to human health and primary energy consumption. Stations also bring tiny contributions, as they represent around 0.06% of each impact.

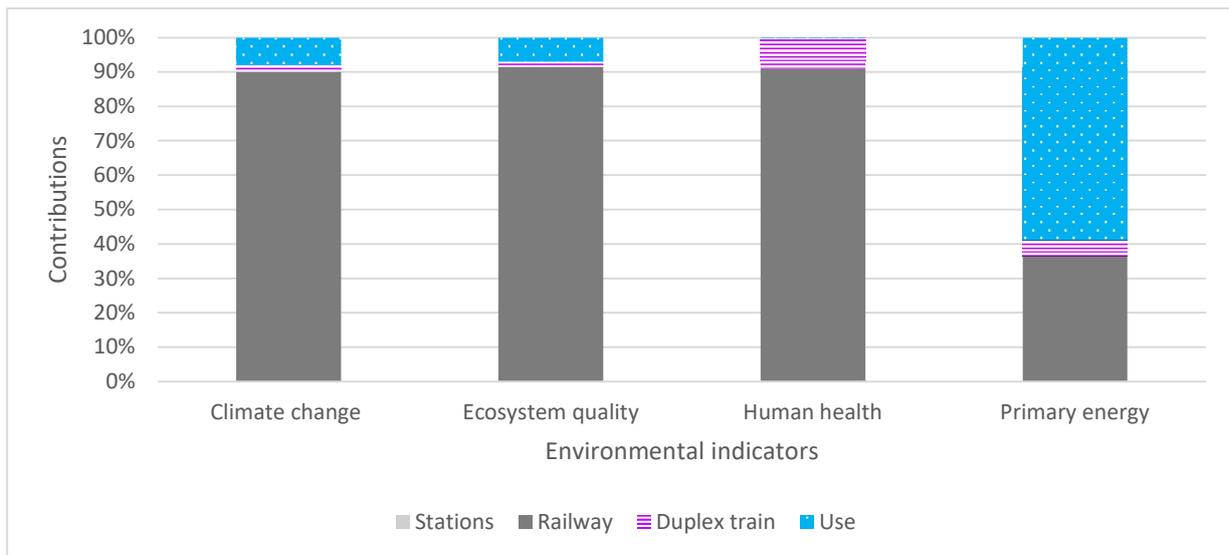

**Figure 4 Contribution of the different components to the total impact for a trip by Duplex train**

4.2.2 Air travel

Figure 5 shows that the use stage has the most important impact for air transportation on the four indicators, followed by the airport, except for human health damage where the trends are opposite. The life cycle contribution to the total impacts of the neo aircraft represents between 0.3 and 4.5%.

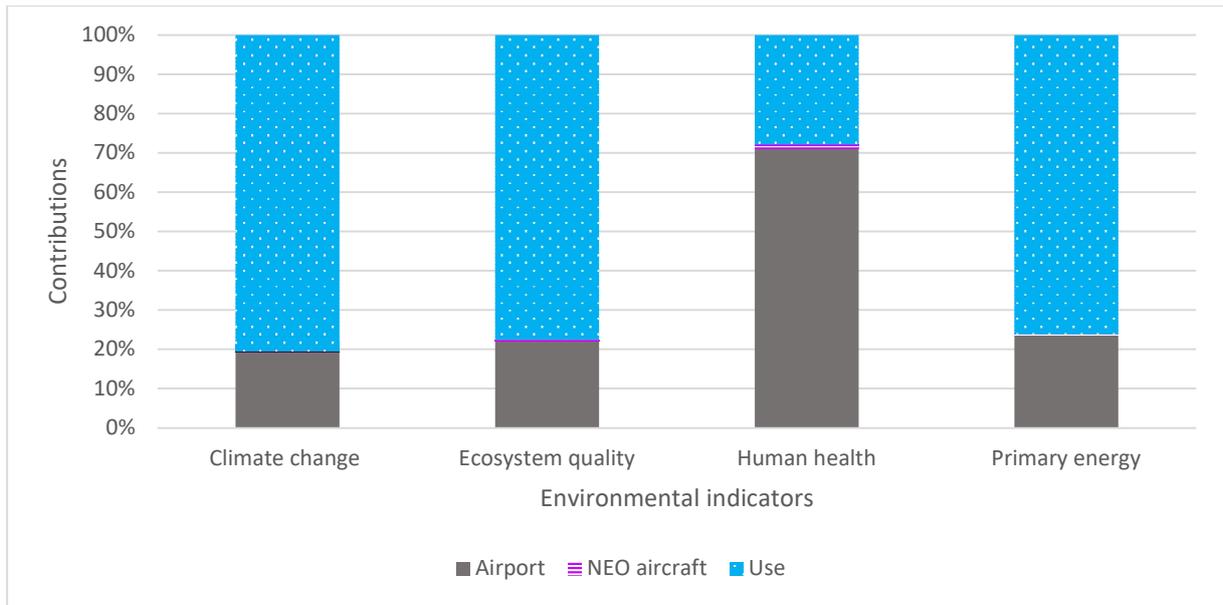

**Figure 5 Contribution of the different components to the total impact when traveling by NEO aircraft**

### 4.2.3 Road travel

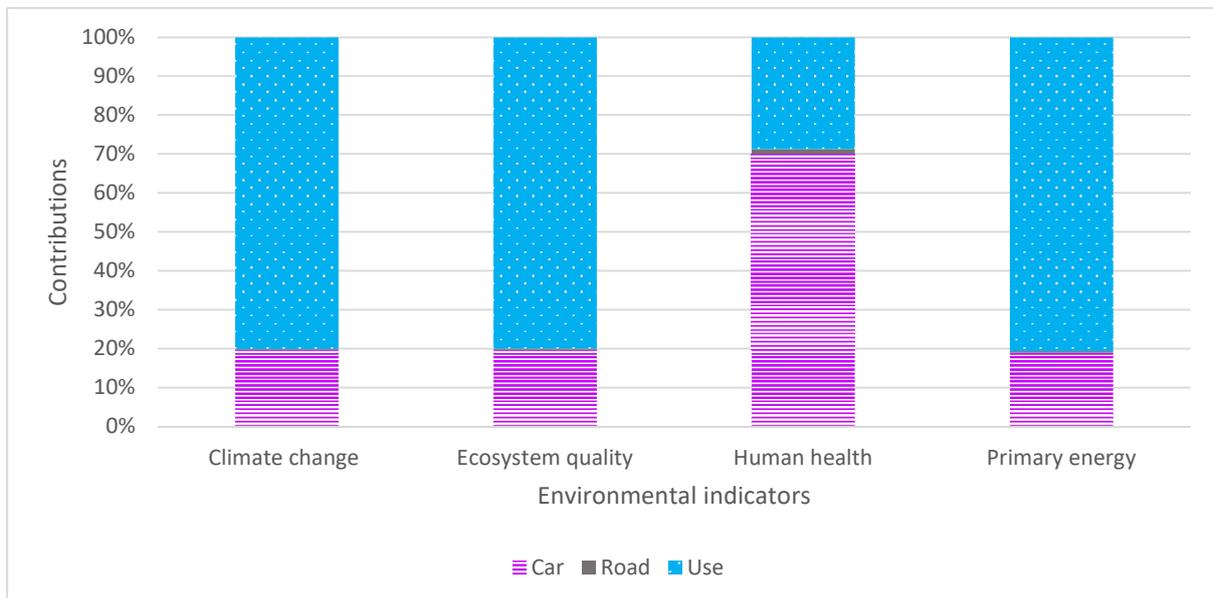

Figure 6 shows that for car mode – personal and carpooling -, the use stage—which includes direct emissions and fuel consumption—is the main contributor for all indicators except for human health again. The vehicle life cycle is the second-biggest contributor, while it is the first contributor for human health due to the use of water. The contribution from pavements is negligible: it represents between 0.1 and 1.5% of each impact. The major and minor contributors are the same for the other road modes modeled in this study. Let's also note that the contribution of the pavement, of course, depends on how infrastructure allocation is performed (see discussion by de Bortoli and Christoforou (2020)).

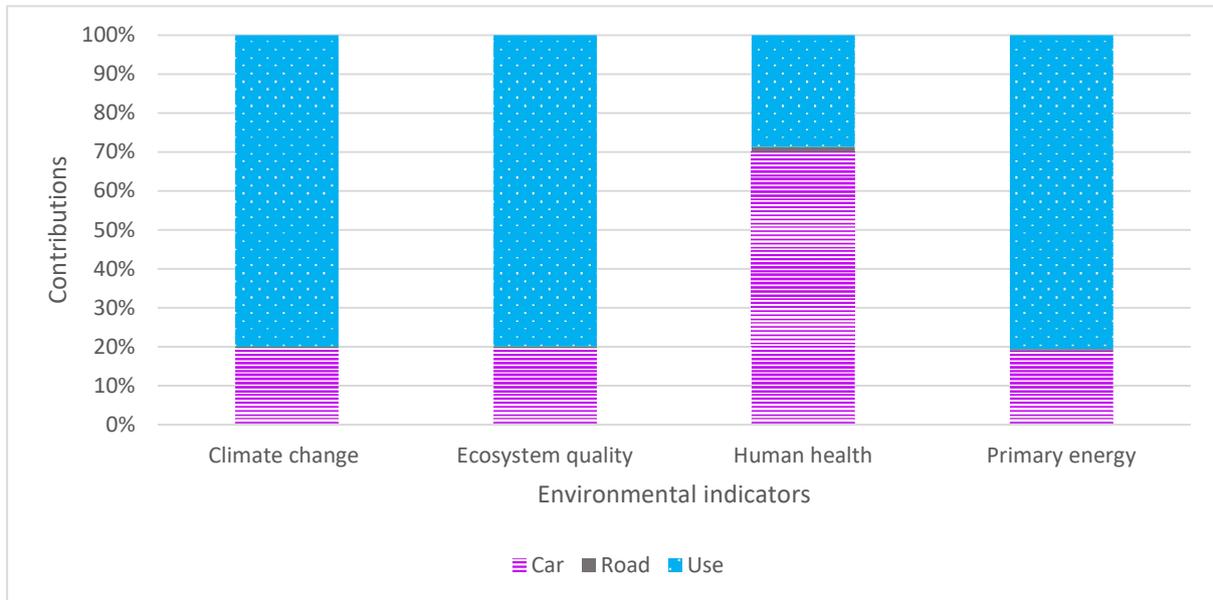

**Figure 6 Contribution of the different components to the total impact for a trip by car (61% diesel, 39% gasoline)**

Figure 7 shows the same trends in contributions from coach modes than from car modes, with a lower contribution from the vehicle manufacturing (except on human health). This is due to a higher intensity of usage for the coach than the car respectively to their manufacturing impacts (5 times higher lifetime mileage and 10 times higher average number of passengers for coaches than cars (e.g. 50:1 functional unit ratio in pkm), for a weight ratio of 10:1).

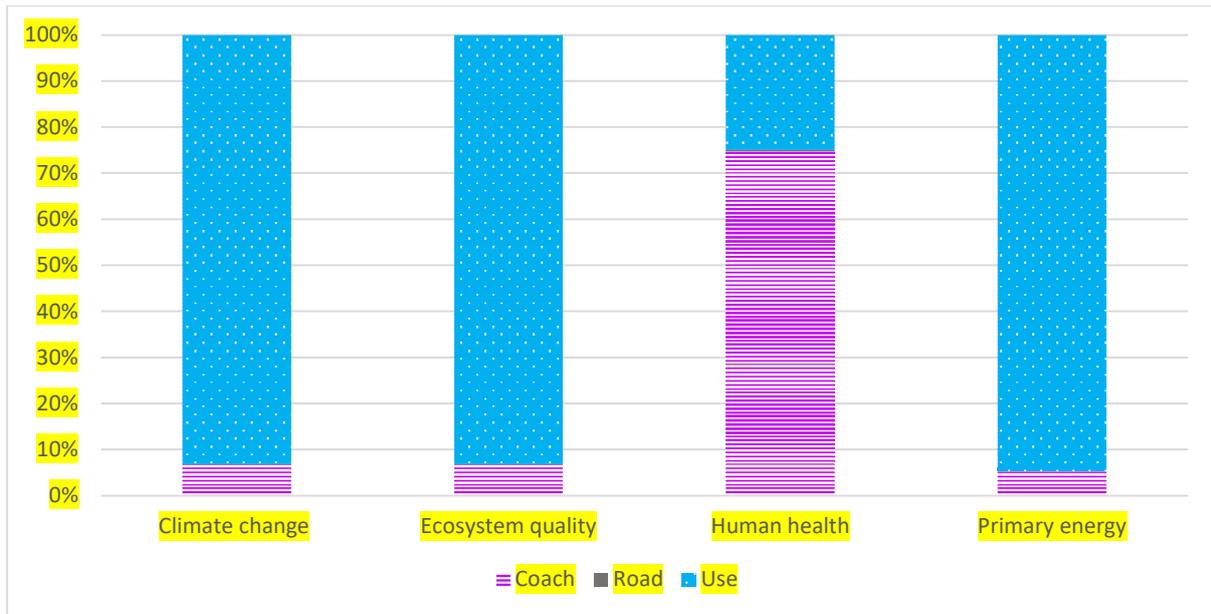

**Figure 7 Contribution of the different components to the total impact for a trip by coach**

### 4.3 ALCA: scenario analyses

4.3.1 Intramodal comparisons: impact of vehicle technology improvement

*4.3.1.1 Rail: Duplex/Atlantique comparison*

Figure 8 presents the comparison of the normalized impacts of train travels (per ptt) using two types of rolling stocks on Paris-Bordeaux: the Atlantique and the Duplex trains. Traveling onboard of a Duplex train presents lower impacts – by 4 to 22% depending on the indicator, and resp. on climate change and human health damage - than traveling onboard of an Atlantique train. This is not explained by the higher energy consumption of the Atlantique rolling stock (27.1 kWh/km compared to 22.9 kWh/km for the Duplex train), but rather by the higher capacity of the Duplex trains (556 versus 454 passengers). Thus, in our case study, the newest rolling stock allows reducing the environmental impact of traveling by train. Nevertheless, the impact in terms of climate change is limited.

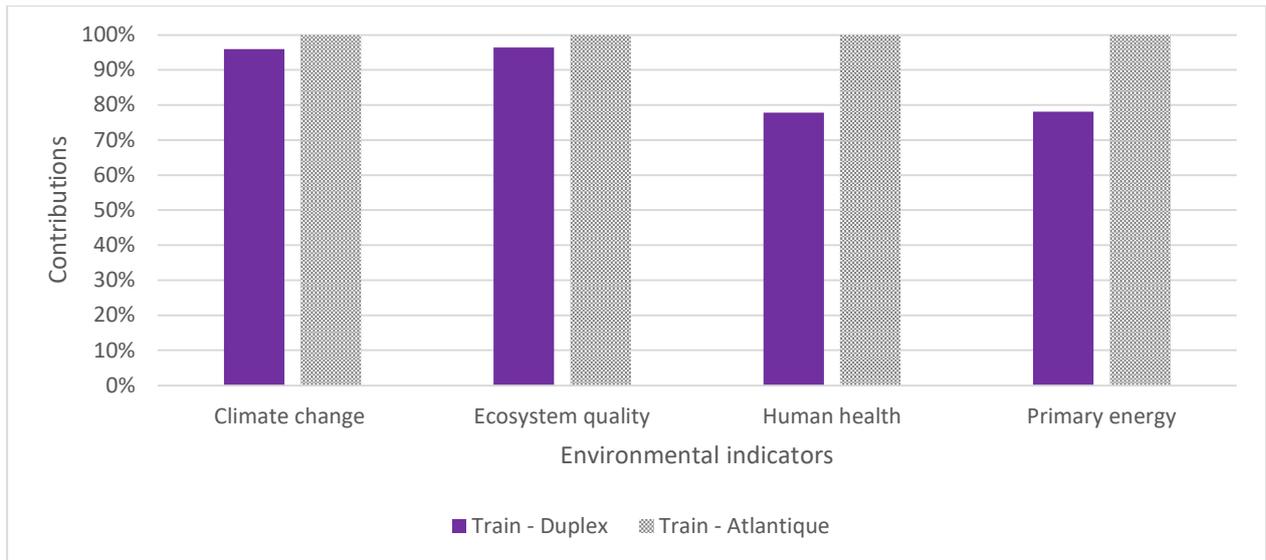

**Figure 8 Train transportation: normalized comparison between a Duplex and an Atlantique carriage**

#### 4.3.1.2 Aircraft type: neo versus ceo comparison

Figure 9 shows the environmental comparison of a Paris-Bordeaux trip, using an A320neo aircraft versus an A320ceo aircraft. While producing an A320neo aircraft is more impacting than manufacturing a ceo version, a trip by neo aircraft presents a better environmental performance on the life cycle by 4 to 24% depending on the indicator, resp. on damage to human health and damage to ecosystem quality. This is due to the use stage, as the neo aircraft has a higher capacity and a better engine performance, thus a lower kerosene consumption. Again, the new technology performs better environmentally.

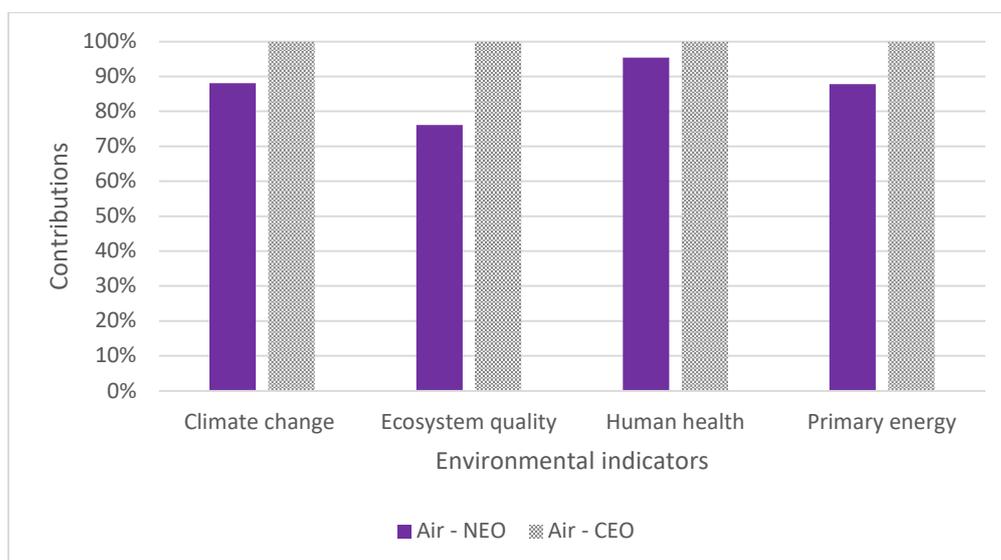

**Figure 9 Air transportation: normalized comparison between CEO and NEO technologies**

*4.3.1.3   Car propulsion: comparison between diesel and gasoline engines*

Figure 10 presents the environmental comparison of a Paris-Bordeaux trip using a diesel car versus a gasoline car. The diesel car has lower impacts—from 7 to 14% depending on the indicator. This can be explained by the use stage which has a lower impact on the diesel car (around 8% less). Indeed, diesel engines consume a smaller quantity of fuel compared to gasoline engines, while the volumetric impact of the two fuels are quite similar. These results based on an LCA are consistent to other studies (e.g. Platt et al., 2017) and opposed to the choice of catch-up of the diesel tax on that of gasoline put in place by the French government on the grounds of increased pollution from diesel engines compared to gasoline engines (Patel, 2013).

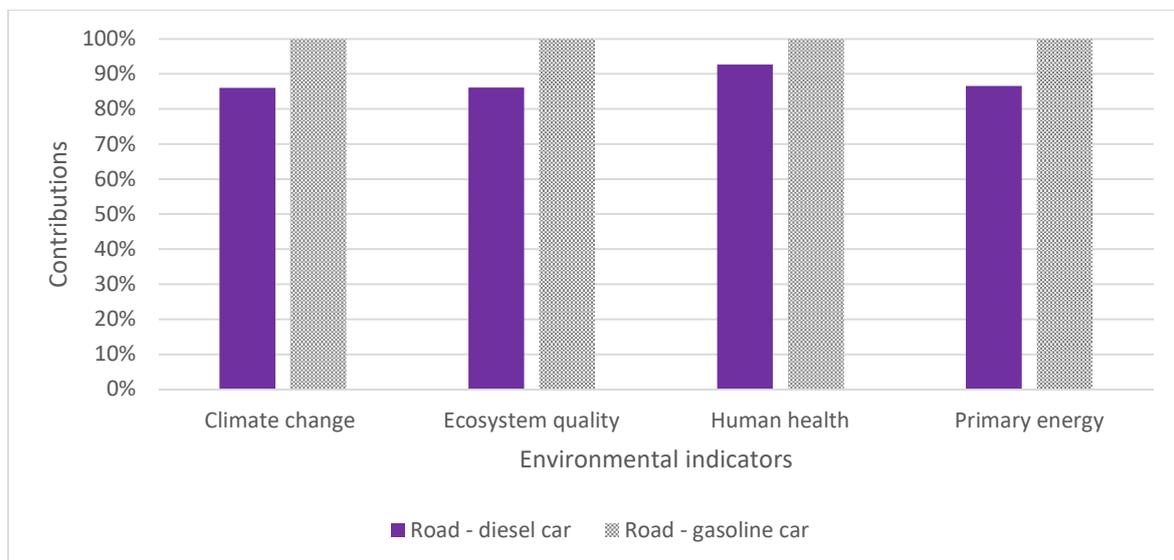

**Figure 10 Car transportation: normalized comparison between diesel and gasoline engines**

4.3.2   Focus on the HSR mode

*4.3.2.1   Electricity mix*

Figure 11 shows the climate change contribution of one pkt by train in different countries. As a simplification, the infrastructure and vehicle are considered identical in the different countries, as well as the occupancy rates, but the electricity mix varies. As steel is the major contributor to the infrastructural impacts, and that it represents a global market, this simplification should be acceptable if rails are reused or recycled in countries with low electricity impacts (to keep recycling impacts minimal). Results show that the climate change contribution depends highly on the electricity mix. The electricity mix with the lower carbon intensity comes from Norway while in Europe, among our selection, the highest intensity can be found in Germany. In this country, the high-speed train mode almost emits 60 $gCO_2eq/pkt$, while the same trip in a medium car (61% diesel, 39% gasoline) emits around 120 $gCO_2eq/pkt$: the HSR is still about 50% less emitting than the average car mode (de Bortoli and Christoforou, 2020). But the mode emits nearly twice more in Germany than in the countries where electricity has the lowest carbon intensity—Switzerland, Norway, and France. Thus, updating electricity mixes before using these figures in other contexts is crucial.

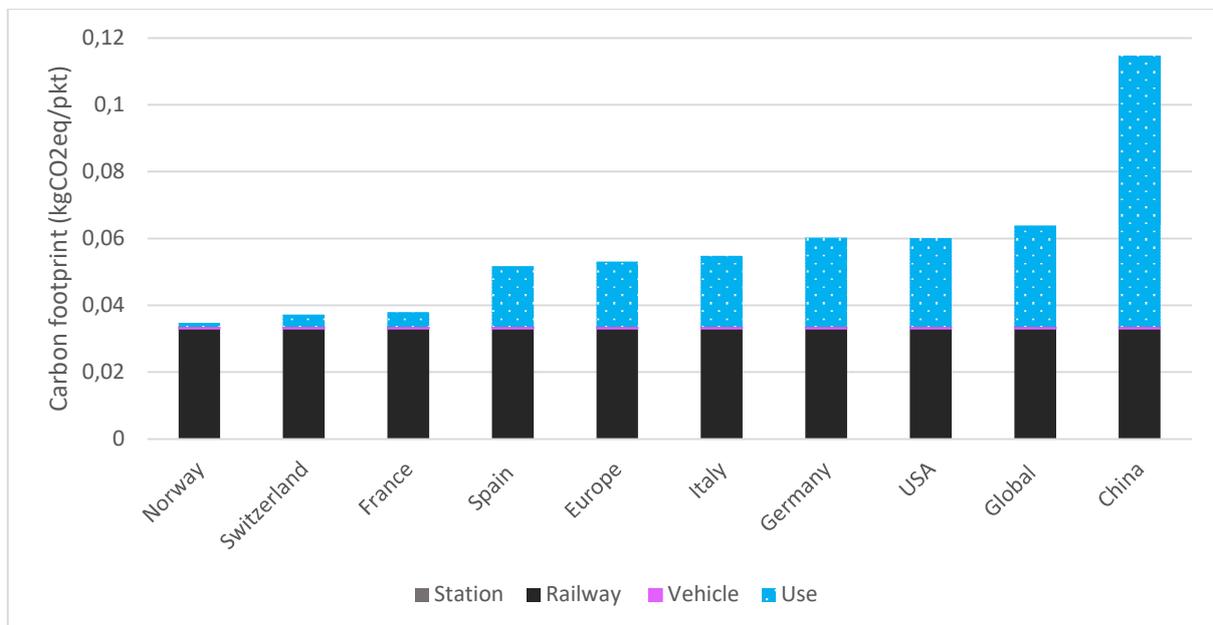

Figure 11 Carbon footprint to travel by duplex HSR in different geographical areas

4.3.2.2    *Optimization of capacities: low-cost offer*

One quarter of the HSR trips between Paris and Bordeaux are bought to Ouigo, the low-cost operator (Provenzano, 2019), with a capacity of the trains jumping from 556 passengers per train in a classic Duplex train to 634 passengers for the low-cost Duplex configuration, by removing the dining car. Moreover, lower fares induce higher demands and finally occupancies: on average, Ouigo trains in France showed an 85% occupancy rate in 2017 according to the general director of the operator company

"voyages SCNF" (Vérier, 2017). A higher occupancy rate means lower environmental impacts per pkt, *ceteris paribus*. According to the French transportation regulation agency, the Ouigo train occupancy rate was 8% higher than the standard train occupancy in 2018 (Autorité de régulation des transports, 2019). On the Paris-Bordeaux line, the Ouigo occupancy rate would thus be around 68 + 8 = 76% in 2018. Figure 12 shows the climate change contribution of one pkt on standard Duplex and low-cost Duplex trains. With the low-cost mode, 12% of $CO_2$ equivalent is saved, falling from almost 38 g$CO_2$eq/pkt to 33 g.

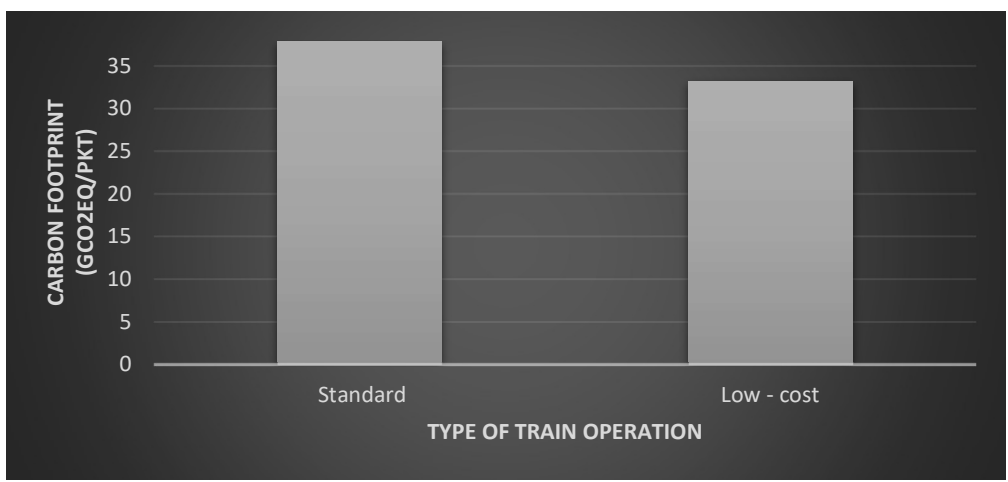

**Figure 12 Carbon footprint to travel by HSR per pkt depending on the type of offer: standard Vs low-cost**

Nevertheless, the induced trips due to attractive fares would account for 40% (Vérier, 2017). This potentially increqses the final environmental impact of the line, by adding passenger trains on the railway, and potentially consuming some railway capacity to the detriment of rail freight, that is more environmentally friendly than their competitors : trucks. Indeed, in France, only 15% of the national railway capacity is used for freight (Autorité de régulation des transports, 2019). The question of railway allocation to reduce overall environmental impacts of transportation, both from freight and passenger transportation, must be addressed holistically, including modal shifts from road and air. But despite the induced traffic due to this low-cost offer, a low-cost offer is socially beneficial as it allows low-income households to afford traveling.

*4.3.2.3   Train's speed*

A classic question to plan railway projects is the choice of the commercial speed. If the trend of recent decades has been to increase the travel speed, we can question the environmental cost of this decision. In fact, in the case of rail, infrastructural constraints—and thus the environmental cost of construction—are higher when speed increases. The maintenance requirements are also higher, as well as the impact of the maintenance. Finally, since the energy consumption of a train would vary according to the square of its speed (Kato et al., 2019), a higher commercial speed means higher environmental impacts too. In the absence of infrastructural data to rigorously compare the impact of two railway projects with different commercial speeds, we will test the sensitivity of the impact to speed based on the operator's data. According to LISEA, the concessionaire of the Tours-Bordeaux HSR, a "TGV"—i.e., a first-generation high-speed train—travels at 230 km per hour in France and consumes 16.5 kWh per kilometer. On the other hand, the new generation high-speed trains travels at 300 km per hour on average, and consumes 22.9 kWh per kilometer. Figure 13 shows that a higher electricity consumption in the case of the new HSR has a negligible impact, increasing by only 4% the carbon footprint per pkt. Nevertheless, in a country with a higher carbon-intensity electricity mix, a higher speed would be more detrimental. It should be noted that, while HSRs are said to be more impacting in terms of construction and maintenance, the environmental impacts for the respective processes in EI "*market for railway track, for a high-speed train, GLO*" and "*market for railway track, RoW*" show that the HSR infrastructure would be less impactful, e.g., emitting almost twice as fewer GHG emissions as the standard railway infrastructure over their life cycles. The EI models might benefit from an update.

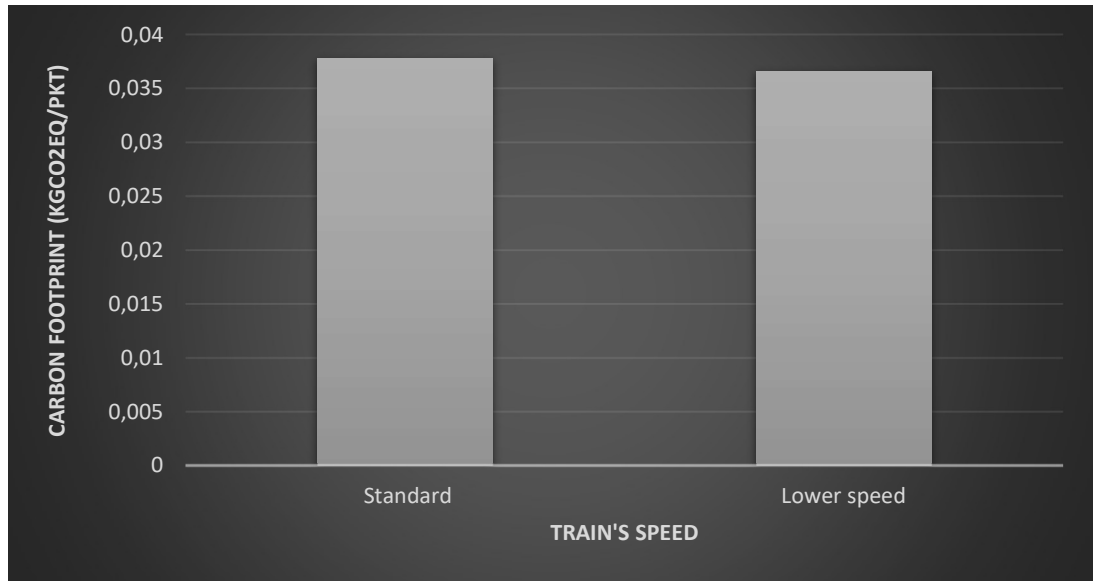

**Figure 13 Carbon footprint per pkt depending on the train's speed: standard (high-speed) vs lower speed.**

### 4.4 CLCA: considering trip substitutions from air to rail

#### 4.4.1 Business-as-usual scenario

Considering a steady annual modal shift, the respective distances between Paris and Bordeaux by plane (507 km) and by rail (497 km), the carbon footprint per pkt with and without the infrastructure resp. by air and rail, and the life-cycle GHG emissions of the HSR infrastructure, we can estimate the time needed for the HSR project to become GHG neutral, i.e., the carbon payback period of the HSR project. We consider the GHG emitted to build the HSR section ($1.4 \cdot 10^9$ kgCO$_2$eq), the carbon footprint of the maintenance over 120 years ($1.1 \cdot 10^9$ kgCO$_2$eq) that we allocate every 30 years except at the EoL, and the benefits from the components' recycling and reuse (-$4.8 \cdot 10^8$ kgCO$_2$eq) that we allocate every 30 years (de Bortoli et al., 2020). Figure 14 shows how the initial carbon investment is repaid twice at the presumed end of life of the railway section (after 120 years). With the modal shifts observed these last few years, the net-zero target (including avoided emissions) would be reached around 60 years after construction. It should be noted that we slightly underestimate the impact of the infrastructure, since we only consider the high-speed section (340 km), the rest of the 497 km being traveled on a conventional rail network that was already built but still need maintenance. But consideration of the expected future reduction in the carbon footprint of the electricity mix in France should on the contrary shorten the carbon payback time. Nevertheless, this new HSR is considered to have made the rest of the rail traffic rise (Arraud, 2019). Whether this rise is due to induced trips or modal shifts from more impacting modes is the key question to understand the environmental consequence of this more complex effect.

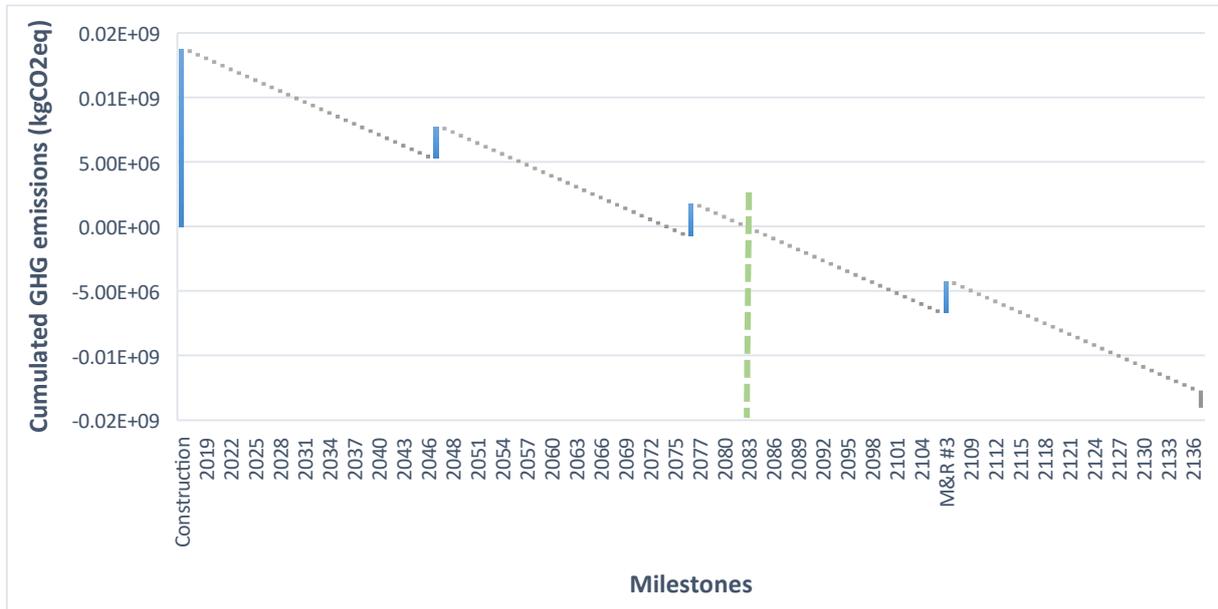

**Figure 14 Net carbon footprint of the HSR project over the years, considering a 200 000-passenger modal shift between planes and rail**

4.4.2 Short-haul flight ban scenario

France voted and published in its "Climate law" at the end of 2021 an article to ban domestic flights for which direct train trips of less than 2 hours and 30 minutes exist (Légifrance, 2021). This is the case on the Paris-Bordeaux route. Later voted by the European commission, if this short-haul flight ban were actually applied in 2022, the GHG emissions due to the HSR construction would be offset thanks to modal shifts within around 10 years (Figure 15). At the end of the life of the HSR, in 2137, the project would have avoided nearly 20 million tons of $CO_2$ equivalent, i.e., the annual emissions of 2.2 million French people considering their current lifestyle (Baude, 2022). These avoided emissions have been estimated considering static life cycle inventories (e.g., non-evolutive electricity mixes), and "first order" modal shifts (i.e., modal shifts from short-haul flights to high-speed train trips exclusively). These aspects should be addressed to refine the calculation. Nevertheless, extending the modal shift calculation to further orders (i.e., evolution in residents' and non-residents' mobility in France and elsewhere due to short-haul flight bans) would necessitate to use traffic simulation models, which are quite uncertain, especially at large scales and in the long run.

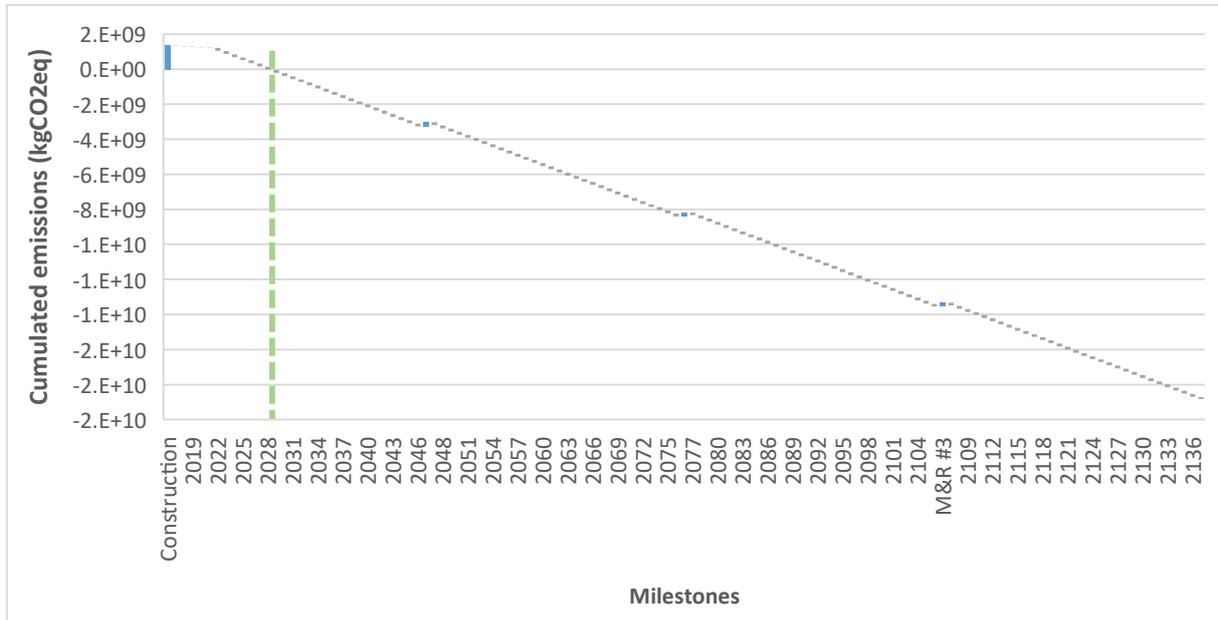

Figure 15 Net carbon footprint of the HSR over the years with a short-haul flight ban

## 5 Discussion

### 5.1 A necessary update of long-distance transportation assessments: comparison with former American LCA estimates

In Figure 16, we compare the carbon footprint and energy consumption of HSR, plane, car, and coach modes on the Paris-Bordeaux corridor with the seminal US results in Chester's work. The HSR performance we consider is an average between "HSR Future WECC-2010 670" and "HSR Future WECC-2010 150" (Chester and Horvath, 2012a) to get approximately the same occupancy rate. The plane considered is the "Bombardier CS-300 ER" with 82 passengers (Chester and Horvath, 2012a) for the same reason. The French gasoline car mode is compared to the US conventional gasoline sedan and the French coach mode with the urban bus on peak for lack of more similar mode (Chester and Horvath, 2009).

In terms of energy consumption, the results are quite similar in France and the US for the HSR modes. Nevertheless, the French HSR mode emits twice as fewer GHG emissions as the US one in 2010. This could have been simply explained by the difference in the electricity mix, but consideration of contributions shows it is more complicated. In the American study, around 50% of the GHG emissions and energy consumption come from the propulsion electricity. In our study, almost 90% of the GHG emissions come from the railway, while 70% of the energy is consumed due to the use stage and 30% due to the railway. The environmental profiles of the HSR mode in the two studies are thus very different. When comparing air modes, the estimates by Chester and Horvath were far lower than those found for Paris-Bordeaux. First, the impact estimated from the airport is higher in our study (50 $gCO_2eq$/pkt) than in the American study (8g), which may be explained by (1) high-traffic airports in the US, (2) infrastructural allocation discrepancy, (3) underestimates in the US model due to the use of EEIO-A instead of process-based LCA, and probable non-endogenized capital goods in the EEIO database used, what would not be suitable to assess airport impacts. The gasoline car in France is less impacting than in the American study, due to 30% lighter cars in France (IEA, 2019) and potentially a higher occupancy considered in the French model. Finally, a high-occupancy urban bus in the US consumes and emits more than a high-occupancy coach in France, maybe due to high consumptions and emissions in urban conditions, while a steady speed at 80 km/h is almost optimal in terms of consumption and thus emissions.

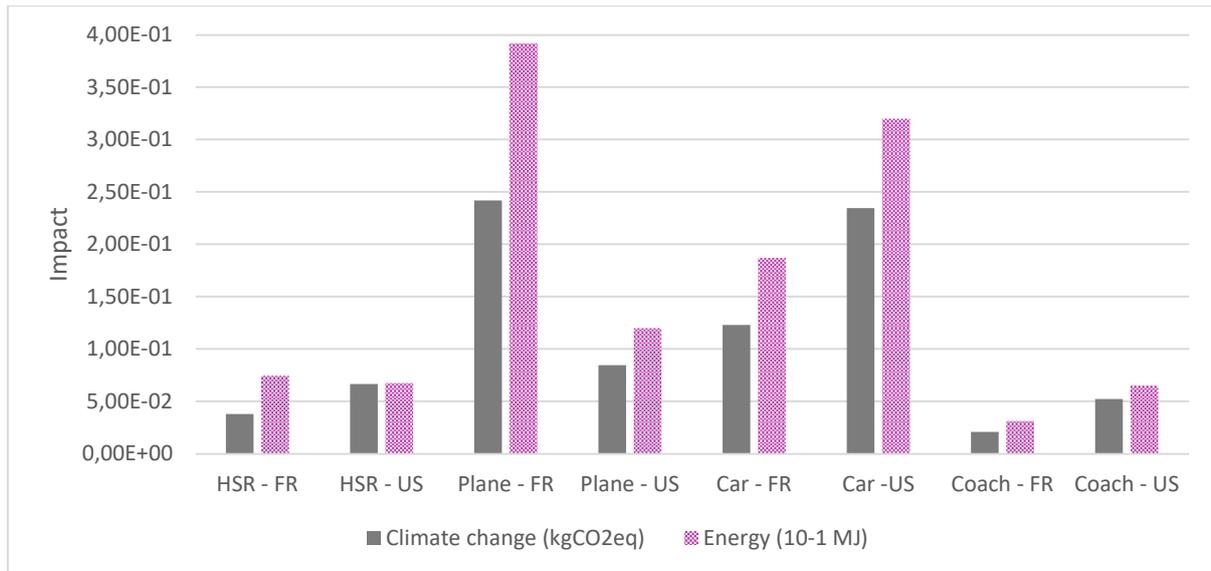

**Figure 16 Comparison of the carbon footprint and energy consumption of HSR, plane, car, and coach modes on the Paris-Bordeaux corridor with the seminal US results by Chester**

Although Chester paved the way for IMLCA, his work was based on a hybrid LCA—i.e., a mix between process-based LCA ("standard" ALCA) and EEIO-A—and partially theoretical data due to lack of field data, bringing some uncertainties in the results. In addition, the geographic relevance of his analysis is restricted to the US or California, and data date back to the 2000s. According to the characterization of data quality used in LCA with the Pedigree matrix (Weidema and Wesnæs, 1996), data older than 15 years is considered to be of the poorest quality, thus potentially generating large uncertainties on the results that do not represent anymore the impacts of current transportation modes. Hence, understanding globally long-distance mobility environmental impacts to compare modes required updated and regionalized LCA models, i.e., a wider geographic and technological coverage. Moreover, most Chester's work uses the seminal US TRACI impact assessment method version, only considering a few environmental dimensions, and currently considered as outdated, for instance due to lack of regionalization of the impacts, outdated GWP, or because considering PM10 instead of PM2.5 that are more harmful to humans. It also does not consider fates of pollutants and exposure effects, will complex causality chains must be considered to understand the final impacts of pollutants on human health or ecosystems (Bulle et al., 2019; Polichetti et al., 2009). Our study thus brings new insights on the environmental pertinence of long-distance travel modes, with indicators focused on biocentric and anthropocentric environmental concerns. To go further and consider technosphere evolutions overtime, future studies should try to include consistent retrospective and prospective background datasets representing the technosphere when activities occur along the transportation system life cycle (de Bortoli et al., 2023). A few prospective attempts have been done on electricity mix evolutions (Robertson, 2016; Wang et al., 2019) and/or rough foreground technology penetration/progress (Åkerman, 2011; Bueno et al., 2017) in HSR LCAs, but they do not consider the global evolution of the global production system. Some first more consistent prospective EI datasets can be calculated using PREMISE (Sacchi et al., 2022). On the other hand, retrospective datasets do not exist, unless somehow the annual versions of EI (de Bortoli et al., 2023). But they do not exactly represent times series, as mostly electricity mixes are updated. Moreover, mixing them may generate practical problems (mapping, non-consideration of errors solved in different versions, allocation evolution, etc.).

## 5.2 Considering the use stage is not enough: environmental transportation policy recommendations must rely on complete life cycle assessments

In France, the NF EN 16258 standard requires the calculation of energy consumed and GHG generated by transportation services, these impacts must be indicated on the services sold (train tickets, plane tickets, freight, etc.): it is the "French carbon label" for transportation services. However, according to the standard, these evaluations must be carried out on scopes 1 and 2: they only consider the use stage of the vehicles. Thus, the national operator indicates that the high-speed train's GHG emissions account for 1.73 $gCO_2eq$/pkt and those of the Ouigo for 0.73 g. Our study highlighted the major contribution of the infrastructure to the carbon footprint of high-speed rail (90%), and thus, that the carbon footprint of the Paris-Bordeaux HSR is 20 higher than the one indicated on the French carbon label. As a conclusion, the NF EN 16258 standard inordinately favors the train to the detriment of other modes in the case of France. Finally, only an evaluation including scopes 1, 2 and 3 according to the GHG Protocol (WBCSD and WRI, 2011)—i.e., on a complete life cycle including the vehicle and infrastructure—will ensure an unbiased environmental comparison of transportation modes and sound policy recommendations.

## 5.3 Generalizing short-haul flight bans and investing in HSR infrastructure to reach sustainable net-zero?

A transportation mode can have highly variable environmental performances, in particular depending on the technology of the vehicles used, their filling rate, the type of energy they consume, and the impact of the infrastructure they use. Thus, saying that one mode is "bad for the environment" as frequently done in popular newspapers, or that one mode is worse than another, is inaccurate. Nevertheless, we can draw some general trends on the performances evaluated in several contexts and with several technologies, as done in this article. At the same occupancy rate, the carbon footprint of ICE cars is relatively stable geographically due to the major impact resulting from direct $CO_2$ emissions. We have shown that the carbon footprint of traveling by HSR highly varies too: our case study and scenario analyses showed a factor of 4, between a little more than 30 $gCO_2$eq/pkt (by combining high occupancies of low-cost offers and low carbon intensity of Norway's electricity mix) and at least 120 $gCO_2$eq/pkt (with an electricity mix mainly based on coal). Overall, this is much lower than emissions due to traveling by personal internal combustion car, but it can compete with carpooling. Besides, we have shown that HSR can also be a decarbonization investment due to the air/rail competition and related modal shifts. An HSR project often generates a spontaneous trip substitution from planes to trains. In this case, the carbon neutrality of the project can be achieved after a relatively long period, such as the 60 years highlighted in our study. On the other hand, with a proactive public policy, the carbon neutrality of a large rail project can be achieved in a few years—less than 10 years in the case of the Tours-Bordeaux rail section with a short-haul flight ban. While the GHG balance will take longer to achieve in countries with high carbon-intensity electricity mixes, rail nevertheless seems to be an investment for a sustainable future to be coupled with adapted transportation policies to reduce the use of fossil fuels and reach net-zero. From a practical point of view, banning short-haul flight when existing acceptable train options are available is partly unpopular, but a feasible policy to maintain fast mobility while respecting Paris' agreement, as air traffic must decrease to reach carbon neutrality (Sacchi, 2023). When no train connection exists, developing new HSR takes several years and can also be unpopular projects due to expropriations and negative externalities for local inhabitants, as well as having strong negative impacts on specific species (habitat losses and reduced accessibility to reproduction areas due to land usage). Thus, new projects must be carefully studied to ensure that modal shifts and HSR traffic would be sufficient to bring global environmental benefits. Otherwise, coach services can perform better than HSR if the vehicles reach a high occupancy rate, like on the Paris-Bordeaux corridor. This questions the affordability of fast long-distance travels on our narrow path to net-zero, and maybe the affordability of any long-distance mobility.

# 6   Conclusions

An integrated transportation LCA model has been developed for the main transportation modes of the Paris-Bordeaux corridor: High-Speed Rail (HSR), plane, and road modes—coach, personal car and carpooling. This model is the first considering properly the complete life cycle of vehicles and infrastructure for each mean of transportation assessed, and especially the crucial rail reuse and recycling for HSR. It also provides new high-quality primary field data and tailored, transparent, open-source models, to minimize uncertainties and contribute to enhancing the community practices usually gathering impact factors erratically from the literature. The ALCA focuses on a large set of environmental indicators focused on two main areas of environmental protection – damages to human health and biodiversity-, as well as on the more common climate change and energy consumption indicators for comparative purposes. ALCA results show that French modes globally rank as follows: coach > HSR > carpooling > private car > plane. Moreover, considering the ptt instead of the standard pkt as the functional unit does not substantially change this ranking, is more accurate, and shows comparative lower impacts of the road modes, as road distances are 15 to 19% higher than other modes' distances. The paper also originally presents the life cycle stage contributions to the environmental impacts for each mode, that can be either the infrastructure, the use stage, or the vehicle amortization, depending on the mode and the indicator considered. Thus, transportation environmental policies must be based on integrated and multi-criteria modal LCAs. Second, several SAs were conducted. They first globally show that technological progress leads to environmental improvements of the transportation modes, from a few percent to 14% for the technologies studied. Also, the low-cost train offer on the Paris-Bordeaux corridor reduces the carbon footprint of the trips by 12% compared to the standard offer, due to a higher occupancy. But changing the train's commercial speed does not impact notably the environmental impact of the mode in France. Moreover, to give an international perspective, we calculated that electric HSR modes would emit between 33 and 120 g$CO_2$eq/pkt depending on the electricity mix considered and the occupancy. Third, considering a consequential approach, we calculated the carbon payback of the Paris-Bordeaux HSR project. Under a business-as-usual scenario, it is reached after 60 years thanks to flight trip substitutions. But with a short-haul flight ban, the carbon payback falls under 10 years. This demonstrates the potential of investing in HSR to decarbonize fast long-distance mobility, as high occupancy coaches are less emitting but much slower. Finally, these figures combined to the carbon paybacks calculated on the Paris-Bordeaux case study advocate for: 1) considering generalizing short-haul flight bans to cut direct emissions from flying that represent most of the impacts of the mode and 2) investing in HSR infrastructure where traffic demand is sufficient, if we decide to maintain massive long-distance mobility in our possible pathways toward net-zero. Otherwise, developing efficient coach services or even reducing long-distance mobility may be better options to decarbonize fast enough our lifestyles to limit dreadful climate change consequences. Finally, as a case study draws some trends but offers limitedly generalizable conclusions - even with sensitivity analyses -, we invite the community to more systematically perform such high-quality, reproducible, and consistent IMLCA, to support decision-making toward a livable future.

**Declarations**

The authors declare that they have no competing interests.

**Availability of data and material**

All the models are available on a repository online, and the excel spreadsheets are provided in the supplementary material, as well as complementary information, to ensure full reproducibility of the study.

**Funding**

None specific funding.

**Authors' credits:** Conceptualization: ADB; Methodology: ADB; Data curation: ADB; Formal analysis: ADB, AF; Software and excel calculations: AF; Validation: ADB; Visualization: ADB, AF; Roles/Writing - original draft: ADB, AF; Writing - review & editing: ADB.

**Acknowledgments**

The genesis of these models comes from the work of two promotions of students (2016–2017 and 2017–2018) from our transportation LCA course at Ponts ParisTech engineering school, financially supported by the Chair ParisTech-VINCI in "Eco-design of buildings and infrastructure". We also thank people from industry who provided field data to us. We thank LISEA, the concessionaire company for the Paris-Bordeaux HSR line, and especially Yannick Depriester, which helped us to collect the data relating to the HSR construction works, the train electricity consumption, traffic date, and updated the maintenance scheme. We thank Yassine Zarrouk for the traffic modeling and the previsions on air and rail traffics. We would also like to thank VINCI Autoroute and especially Cécile Giacobi for the data on the highway A10. Moreover, our thanks go to the experts from Alstom for the data on train manufacturing and life cycle. Finally, a big thank to Dr Maxime Agez for reading the manuscript, his interpretation on EEIO and process-based results discrepancy, as well as his help on understanding results mistakes from AWARE using ecoinvent in OpenLCA.

# Supplementary material - Banning short-haul flights and investing in high-speed railways for a sustainable future?

Anne de Bortoli[1,2*], Adélaïde Feraille[3]
**1** *LVMT, Ecole des Ponts ParisTech, University Gustave Eiffel*
**2** *CIRAIG, Chemical engineering department, Polytechnique Montreal*
**3** *Lab. Navier Laboratory, Ecole des Ponts ParisTech, University Gustave Eiffel*
**\*** Corresponding author; e-mail: anne.de-bortoli@enpc.fr

**Table of contents**

## 6.1 List of tables

## 6.2 List of figures

***Nota Bene:*** *damage to human health from electricity consumption have been recalculated with ei 3.6 on Simapro, using the same IW+ method, as we noticed that the results given with OpenLCA with ei 3.2 and IW+ where wrong due to uncorrect assessment on water scarcity related to hydroelectricity.*

# 7  1. Complementary LCIs

**Table 3 HSR civil engineering structures and material requirement**

| CONCRETE MATERIALS | CONCRETE C30/37 | CONCRETE 30-32 MPA - GLO | 3.46E+04 | M3 |
|---|---|---|---|---|
| | concrete C25/30 | concrete 25MPa - GLO | 3.64E+04 | m3 |
| | concrete C20/25 | concrete 20 MPa - GLO | 2.83E+03 | m3 |
| | concrete C35/45 | concrete 35 MPa - GLO | 1.54E+05 | m3 |
| | concrete C40/50 | concrete 50MPa - GLO | 2.59E+04 | m3 |
| REINFORCED STEEL | reinforced steel | Market for reinforcing steel production - GLO | 7.14E+04 | t |
| WATER SEALING | bitumen-polymer | bitumen seal, polymer EP4 flame retardant - RER | 4.60E+02 | t |
| CORNICES | concrete | concrete, normal - GLO | 1.05E+04 | m3 |
| | reinforced steel | Market for reinforcing steel production - GLO | 8.98E+02 | t |
| FOUNDATION | medium capacity stakes | concrete high exacting requirement, GLO | 7.68E+03 | m3 |
| | high capacity stakes | concrete high exacting requirement, GLO | 4.17E+04 | m3 |
| STEEL STRUCTURE | steel | Market for reinforcing steel production - GLO | 7.90E+03 | t |

**Table 4 HSR maintenance scheme**

| COMPONENT | OPERATION | PERIOD (YEAR) | PART | EOL |
|---|---|---|---|---|
| RAIL | Milling | 1 | 100% | N/A |

|  | Replacement | 40 | 100% | 80% recycling - 20% reuse |
|---|---|---|---|---|
| **BALLAST** | Tamping | 1 | 50% | N/A |
|  | Backfilling | 20 | 15 cm | Recycling |
|  | Replacement | 30 | 30% | Recycling |
| **CHAIRS** | Replacement | 30 | 100% | Landfill disposal |
| **FASTENERS** | Replacement | 30 | 100% | Recycling |
| **SLEEPERS** | Replacement | 60 | 100% | Recycling |

**TABLE 5** Materials and energy consumption for the construction of the power supply and signaling system -EcoInvent processes chosen in the model

| PROCESS DETAIL | ECOINVENT PROCESS | QUANTITY | UNIT |
|---|---|---|---|
| *TRENCHES* | | | |
| EARTHWORK AND BUILDING MACHINES | Diesel, burned in building machine - GLO | 23115894 | MJ |
| CONCRETE | concrete, normal - GLO | 4046.90022 | t |
| REINFORCED STEEL | steel, converter, low alloyed - RER | 2782.2439 | t |
|  | hot rolling, steel - RER | 2782.2439 | t |
| TRANSPORTATION | transport, freight, lorry 16-32 metric ton, EURO5 | 8.35E+05 | tkm |
| *CATENARY CABLES* | | | |
| CONDUCTING CABLES | cooper - RER | 1.67E+06 | kg |
|  | wire drawing, copper - RER | 1.67E+06 | kg |
|  | steel, low-alloyed, converter - RER | 1.28E+05 | kg |
|  | hot rolling, steel - RER | 1.28E+05 | kg |
|  | market for bronze production - GLO | 1.05E+06 | kg |
| SUPPORTING CABLES | aluminum, cast alloy - RER | 9.75E+05 | kg |
|  | Market for sheet rolling, aluminun - GLO | 9.75E+05 | kg |
| TRANSPORTATION | transport, freight, lorry 16-32 metric ton, EURO5 | 1.14E+06 | tkm |

# 8  2. Details on the method

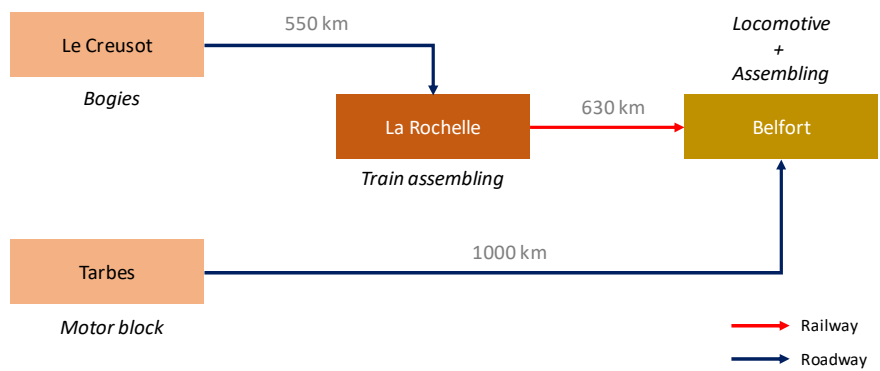

**Figure 17** Transportation scheme for the train construction

**Table 6 Airframe transportation details**

| Trip | Broughton → Toulouse | Saint-Nazaire → Toulouse | Hamburg → Toulouse |
|---|---|---|---|
| **Transportation mode** | Truck | Truck | Belugas (aircraft) |
| **Distance** | 1225 km | 591 km | 1500 km |
| **Component** | Wings | Forward-fuselage | Central fuselage and tail |
| **Total weight contribution (%)** | 25 | 25 | 50 |

# 9   3. Aerial traffic evolution on Paris-Bordeaux

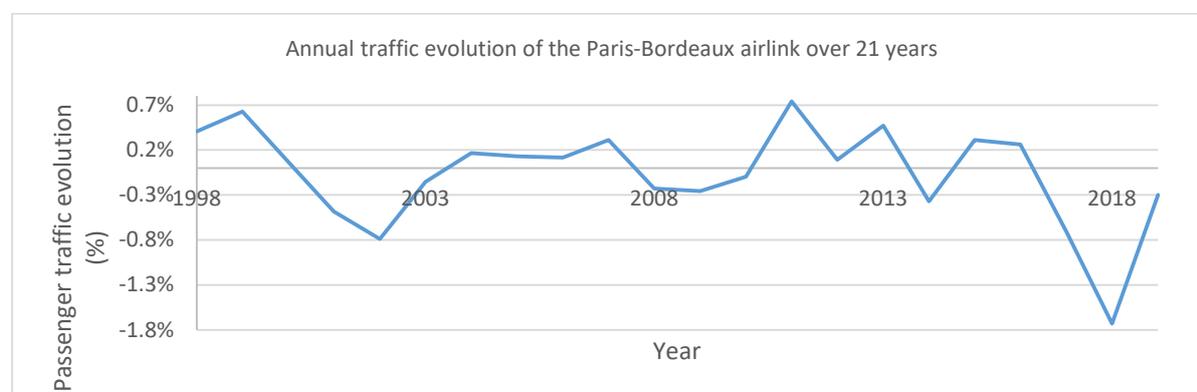

**Figure 18 Annual traffic evolution of the Paris-Bordeaux airlink between 1998 and 2019 over 21 years (%) (Data: Annual DGAC figures)**

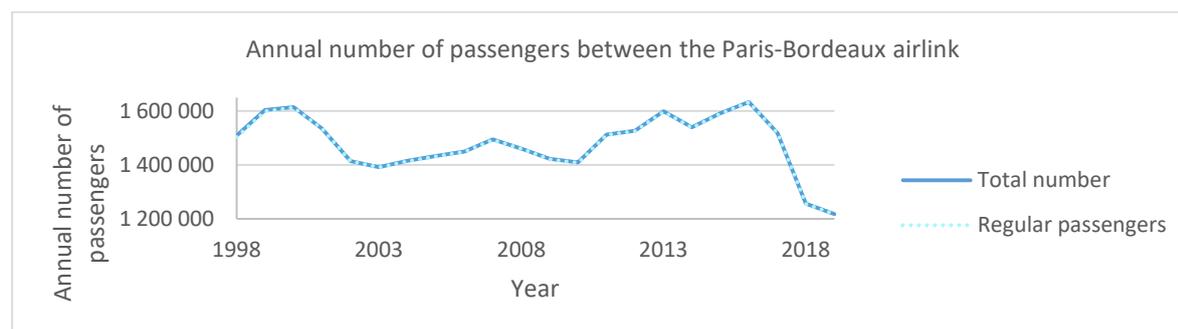

**Figure 19 Annual number of passengers on the Paris-Bordeaux airlink between 1998 and 2019 (Data: Annual DGAC figures)**

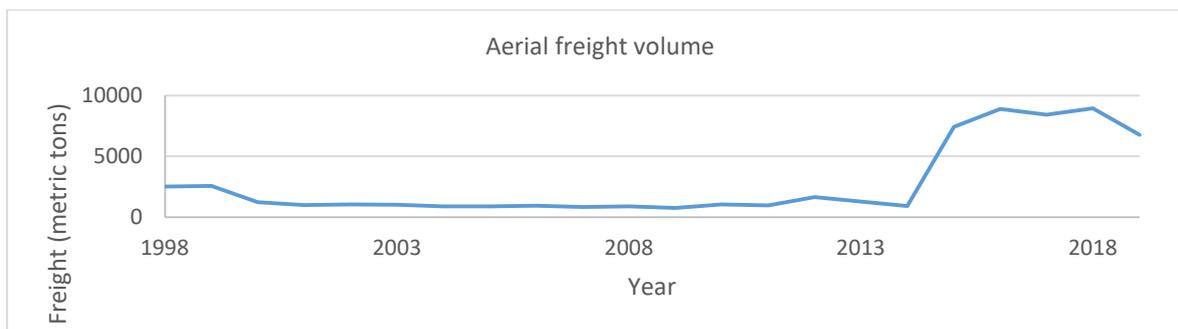

**Figure 20 Annual aerial freight volume over Paris-Bordeaux between 1998 and 2019 (Data: Annual DGAC figures)**

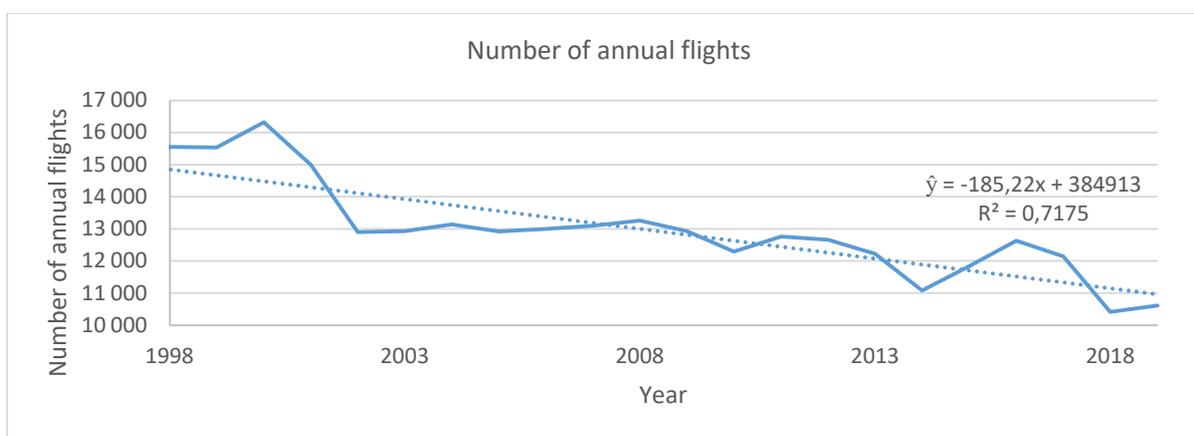

**Figure 21 Number of annual flights on the Paris-Bordeaux airlink between 1998 and 2019 and evolution trend in dotted line (Data: Annual DGAC figures)**

# 10  4. Complementary results and graphs

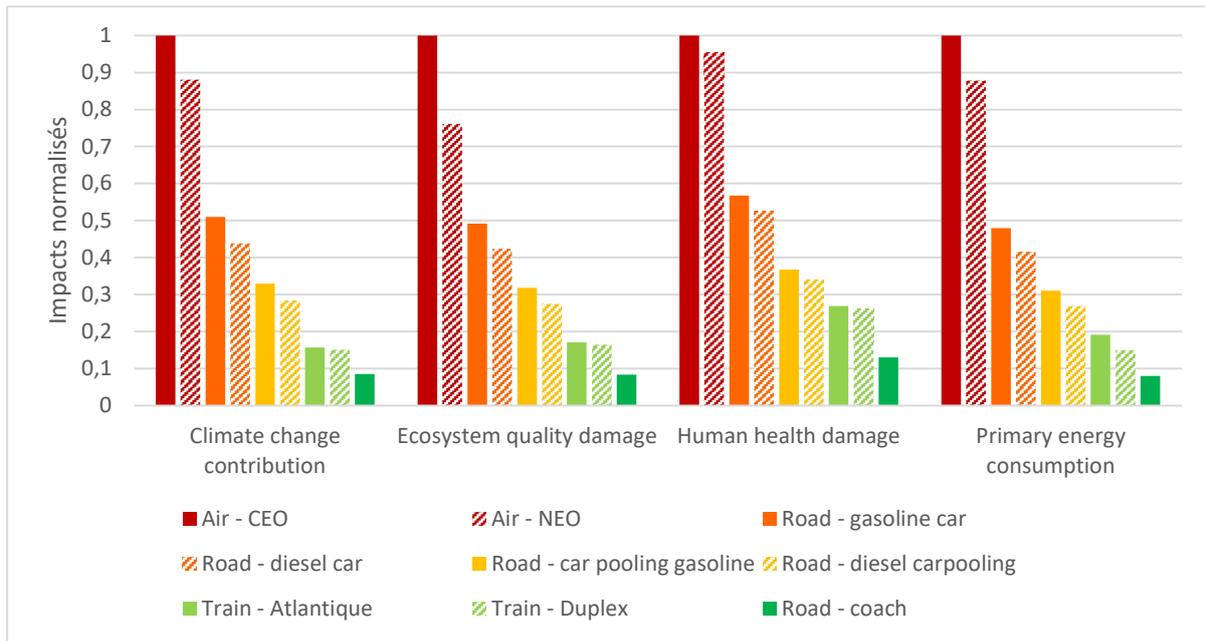

**Figure 22 Environmental ranking of the different modal alternatives on Paris-Bordeaux's link**

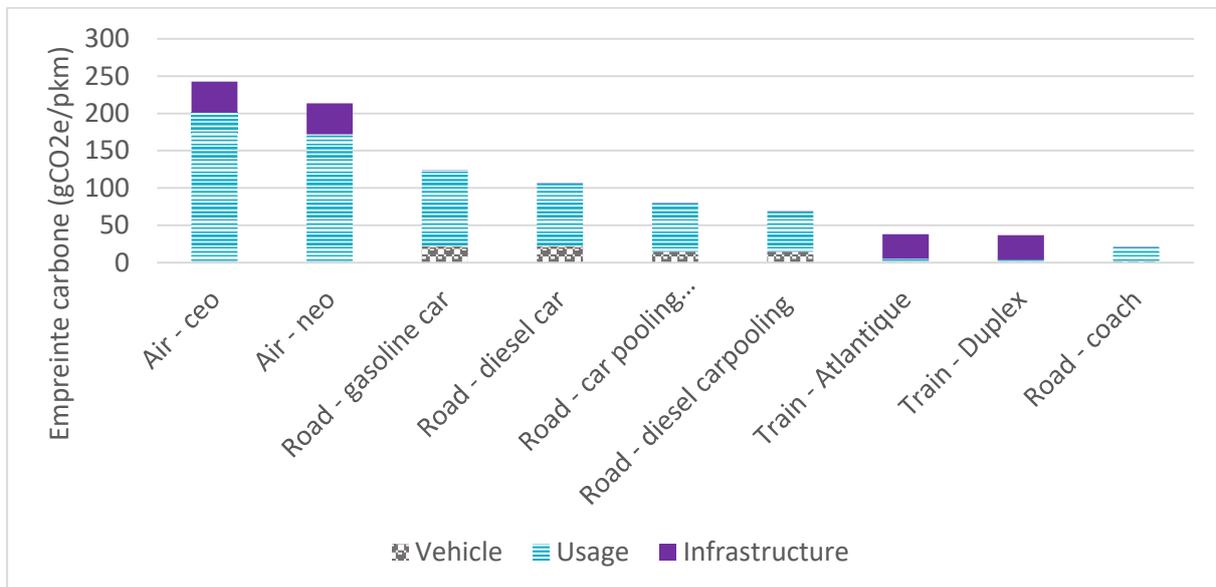

**Figure 23 Carbon footprint and their compposition by mode on Paris-Bordeaux**